\documentclass[11pt,a4paper]{article}

\usepackage[utf8]{inputenc}
\usepackage[T1]{fontenc}
\usepackage{lmodern}
\usepackage[english]{babel}

\usepackage[a4paper,left=2.5cm,right=2.5cm,top=2.8cm,bottom=2.8cm]{geometry}
\usepackage{setspace}
\usepackage{indentfirst}        
\usepackage{microtype}

\makeatletter
\renewenvironment{abstract}%
  {\begin{center}%
     {\bfseries\abstractname\vspace{-.5em}\vspace{\z@}}%
   \end{center}%
   \quotation}
  {\endquotation}
\makeatother

\usepackage{amsmath,amssymb,amsfonts,amsthm,mathtools}
\usepackage{bm}
\IfFileExists{siunitx.sty}{%
  \usepackage{siunitx}
  \sisetup{detect-all,per-mode=symbol}
}{}

\usepackage{graphicx}
\usepackage{float}
\usepackage{booktabs}
\usepackage{longtable}
\usepackage{multirow}
\usepackage{array}
\usepackage{caption}
\usepackage{subcaption}
\usepackage{tikz}
\usetikzlibrary{arrows.meta,positioning,calc,shapes.geometric,plotmarks}
\usepackage{pgfplots}
\pgfplotsset{compat=1.18}

\IfFileExists{algorithm.sty}{\usepackage{algorithm}}{}
\IfFileExists{algpseudocode.sty}{\usepackage{algpseudocode}}{}
\usepackage{listings}
\usepackage{xcolor}
\usepackage{enumitem}
\setlist{itemsep=2pt,topsep=3pt}
\usepackage[numbers,sort&compress]{natbib}
\usepackage[hidelinks,colorlinks=true,
            linkcolor=blue!60!black,
            citecolor=green!50!black,
            urlcolor=blue]{hyperref}
\usepackage{cleveref}   

\theoremstyle{definition}

\theoremstyle{remark}

\newcommand{\dd}{\mathrm{d}}
\newcommand{\ee}{\mathrm{e}}

\newcommand{\defname}[1]{\textbf{#1}}

\title{Six-Degree-of-Freedom Rotor-Load Cross-Spectral Density from
Rotor-Plane Turbulence Cross-Spectra}
\author{Haoran Yang\thanks{Corresponding author: \texttt{haoranyang\_12@sjtu.edu.cn}.} \\
        \small School of Ocean and Civil Enginerring / Shanghai Jiao Tong University
        \and
        Zhaolong Han \thanks{Corresponding author: \texttt{han.arkey@sjtu.edu.cn}} \\
        \small School of Ocean and Civil Enginerring / Shanghai Jiao Tong University}

\date{2026-10-23}
\begin{document}
\maketitle


\begin{abstract}
A frequency-domain response analysis of a wind turbine rotor requires the complete
hub-load spectrum, but the customary scalar thrust and torque spectra retain neither
the four in-plane components nor the coupling between them. This paper derives the
full six-degree-of-freedom (6-DOF) rotor-load cross-spectral-density (CSD) matrix
from the IEC Kaimal turbulence model. By decoupling the spatial order of inflow
coherence from the temporal Doppler shift, the formulation introduces a two-index
rotational-sampling kernel. Rotor symmetry and azimuthal averaging constrain the
observed Doppler carrier strictly to blade-passing harmonics $pBf_{\mathrm{rot}}$,
while the azimuthal aerodynamic transfer shifts the spatial coherence order to
$pB+s$. A convergence study of the retained azimuthal order $s$ shows that
$\lvert s\rvert\le4$ suffices, the assembled CSD then deviating from the full-band
result by less than $10^{-5}$ in relative norm. The formulation is compared with
OpenFAST simulations of a rigid floating wind turbine over nine wind speeds: the
six auto-spectral densities agree closely below 1~Hz, and the present theory
over-predicts the spectra near the 6P and 9P rotational harmonics.
\end{abstract}

\clearpage
\tableofcontents
\clearpage


\section*{Notation}

\addcontentsline{toc}{section}{Notation}


The rotor load vector is expressed at the hub center in a hub-fixed rotor frame $R$:

\begin{equation}\label{eq:rotor-load-vector}
  \delta\boldsymbol{L}=
  \begin{bmatrix}
  \delta F_x & \delta F_y & \delta F_z &
  \delta Q_x & \delta Q_y & \delta Q_z
  \end{bmatrix}^{T}\in\mathbb{R}^6 .
\end{equation}

\begin{itemize}[leftmargin=*]
  \item $\boldsymbol{L}$ is the six-component hub-load vector
        and $\delta\boldsymbol{L}$
        its perturbation;
        $a,d\in\mathcal{L}$ label its components,
        with $\mathcal{L}=\{F_x,F_y,F_z,Q_x,Q_y,Q_z\}$.
  \item $k\in\{u,v,w\}$ denotes a wind-aligned turbulence component.
  \item $B$ is the number of blades,
        $N_r$ is the number of radial elements per blade,
        and $N_\psi$ is the number of reference azimuths used for azimuthal averaging
        or for stationary CSD averaging.
  \item $b,c\in\{1,\ldots,B\}$ are blade indices: blade $b$ carries the azimuth offset
        $\beta_b=2\pi(b-1)/B$ and the azimuth $\psi_b(t)$, with $\psi_0$ the initial
        rotor orientation and $\Omega$ the rotor speed.
  \item $i,j\in\{1,\ldots,N_r\}$ are radial-station indices with radii $r_i$, $r_j$,
        and $u_{ib}(t)$ is the longitudinal inflow perturbation at station $i$ on
        blade $b$.
  \item $t$ is time, $\tau$ is a time lag, $f$ is the observed frequency, $\nu$ is the
        fixed-frame turbulence frequency, and $f_{\mathrm{rot}}=\Omega/(2\pi)$ is the
        rotor frequency.
  \item $\kappa\in\{1,\ldots,N_\psi\}$ is the azimuth-sample index, with $\psi_\kappa$
        the $\kappa$th sampled azimuth; $\mu$ and $\ell$ index the azimuthal quadrature
        nodes $\theta_\mu$ and the output-frequency bins $f_\ell$ of the discrete
        evaluation in \cref{sec:numerical}.
  \item $\gamma(f,d)$ is the spectral coherence of two rotor-disk points a distance
        $d$ apart, and $S_u(f)$ is the longitudinal turbulence spectrum.
  \item $F_n(f;r_i,r_j)$ is the $n$th azimuthal Fourier coefficient of the coherence
        on the station pair $(i,j)$, that is, the $n$th coherence harmonic that
        enters the rotational-sampling kernel.
  \item $g_i^{(a)}(\psi)$ is the load-transfer coefficient of component $a$ at
        station $i$, as a function of rotor azimuth $\psi$, and $\hat{g}_{i,s}^{(a)}$
        is its $s$th azimuthal Fourier coefficient.
  \item $n$ is the spatial Fourier order of the rotor-plane coherence, $s$ is the
        azimuthal Fourier order of the load-transfer coefficient, $q=n-s$ is the
        temporal frequency-shift order of the resulting hub load, and $p$ is the
        blade-passing order of the load spectral families.
  \item $K_{ij}^{(n,q)}(f)$ is the two-index rotational-sampling kernel. Its first
        index selects $F_n$, while its second index sets the shifted frequency
        $f-qf_{\mathrm{rot}}$.
  \item $R_{ij}^{bc}(\tau)$ is the fixed-frame wind cross-covariance between stations
        $(i,b)$ and $(j,c)$; $R_{ad}(t,\tau)$ is the corresponding hub-load
        cross-covariance, and $\overline{R}_{ad}(\tau)$ its azimuthal average.
  \item $\mathbf{S}_{FF}^{\mathrm{rotor}}\in\mathbb{C}^{6\times6}$ is the rotor-load CSD.
\end{itemize}

\newpage

\section{Preliminaries}\label{sec:preliminaries}


A scalar auto-spectrum $S_{xx}(f)$ pertains to a single process at a single
point: it retains the magnitude of that process and no phase information
relative to any other.
The rotor problem, however,
involves $n_{\mathrm{c}}$ spatially separated processes whose
fluctuations are mutually correlated,
so the scalar description must be lifted to a
matrix description.
Let $\bm{x}(t)$, a generic zero-mean $n_{\mathrm{c}}$-dimensional stochastic process,
be arranged as a column vector
\begin{equation}\label{eq:channel-vector}
  \bm{x}(t)=
  \begin{bmatrix}x_1(t) & x_2(t) & \cdots & x_{n_{\mathrm{c}}}(t)\end{bmatrix}^{\rm{T}}
  \in\mathbb{C}^{n_{\mathrm{c}}\times1}.
\end{equation}
The \defname{Covariance Matrix} is defined as the expectation of the
outer product of the column vector $\bm{x}(t)$
with the conjugate transpose of its forward time shift
$\bm{x}^{\rm{H}}(t+\tau)$,
\begin{equation}\label{eq:covariance-matrix}
  \mathbf{R}_{xx}(\tau):=\mathbb{E}\!
  \left[
    \bm{x}(t)\bm{x}^{\rm{H}}(t+\tau)
    \right]
  \in\mathbb{C}^{n_{\mathrm{c}}\times n_{\mathrm{c}}}.
\end{equation}
Its diagonal entries are the scalar auto-covariances and
its off-diagonal entries are the cross-covariances.
Based on the Wiener-Khinchin theorem, the
\defname{Cross-Spectral Density (CSD)} matrix is defined
as the Fourier transform pair of
\cref{eq:covariance-matrix},
\begin{equation}\label{eq:csd-definition}
\mathbf{S}_{xx}(f)=
\int_{-\infty}^{\infty}\mathbf{R}_{xx}(\tau)e^{-\mathrm{i}2\pi f\tau}\,\mathrm{d}\tau,
\qquad
\mathbf{R}_{xx}(\tau)=
\int_{-\infty}^{\infty}\mathbf{S}_{xx}(f)e^{\mathrm{i}2\pi f\tau}\,\mathrm{d}f,
\end{equation}
and $\mathbf{S}_{xx}(f)\in\mathbb{C}^{n_{\mathrm{c}}\times n_{\mathrm{c}}}$ is the double-sided CSD matrix.
Element by element, \cref{eq:csd-definition} gives the auto-spectra on the diagonal,
\begin{equation}\label{eq:diagonal-psd}
  S_{ii}(f)=\int_{-\infty}^{\infty}R_{ii}(\tau)e^{-\mathrm{i}2\pi f\tau}\,\mathrm{d}\tau,
  \qquad i\in\{1,\ldots,n_{\mathrm{c}}\},
\end{equation}
which are the scalar PSDs, and the cross-spectra off the
diagonal. The strength of the coupling between two processes is measured by the
\defname{Coherence},
the modulus of the cross-spectrum normalized by the two
auto-spectra,
\begin{equation}\label{eq:coherence-definition}
  \gamma_{ij}(f)\coloneqq\frac{\lvert S_{ij}(f)\rvert}{\sqrt{S_{ii}(f)S_{jj}(f)}},
  \qquad 0\le\gamma_{ij}(f)\le1 .
\end{equation}
The bound follows from the Cauchy-Schwarz inequality, so $\gamma_{ij}$ is a
dimensionless correlation coefficient: $\gamma_{ij}=1$ for perfectly correlated
processes and $\gamma_{ij}=0$ for uncorrelated ones.
The off-diagonal entry is then recovered from the two
auto-spectra and the coherence in polar form,
\begin{equation}\label{eq:off-diagonal-csd}
  S_{ij}(f)=\gamma_{ij}(f)\sqrt{S_{ii}(f)S_{jj}(f)}\,e^{\mathrm{i}\varphi_{ij}(f)},
\end{equation}
where $\varphi_{ij}(f)=\arg S_{ij}(f)$ is the phase of the complex cross-spectrum.
\Cref{eq:off-diagonal-csd} is the polar form of \cref{eq:csd-definition} and fixes no
cross-spectrum by itself; a coherence model is required, as introduced for the
rotor-plane processes in \cref{sec:kaimal}.


\section{Rotor Excitation Spectrum: Complete 6-DOF Formulation}\label{sec:rotor-csd}


Of the three turbulence components $k\in\{u,v,w\}$, only the longitudinal one $u$ is
retained: it sets the relative inflow speed and hence the angle of attack of every
blade section, so it dominates the unsteady rotor aerodynamics.
For an ideal rotor with the wake induction neglected, the relative inflow at radial
station $r$ has the axial component $U_{\mathrm{hub}}$ and the tangential component
$\Omega r$, so the inflow angle $\varphi=\arctan x$ is fixed by
$x=U_{\mathrm{hub}}/(\Omega r)$.
The axial perturbation $\delta u$ varies the numerator of $x$ and the tangential
perturbation $\delta v_t$ its denominator; with $\dd(\arctan x)=\dd x/(1+x^{2})$, the
first variation is
\begin{equation}\label{eq:inflow-angle-perturbation}
  \delta\varphi
  =
  \frac{\delta\!\left(U_{\mathrm{hub}}/(\Omega r)\right)}
       {1+\left(U_{\mathrm{hub}}/(\Omega r)\right)^{2}}
  =
  \frac{\left(\Omega r\,\delta u-U_{\mathrm{hub}}\,\delta v_t\right)/(\Omega r)^{2}}
       {1+U_{\mathrm{hub}}^{2}/(\Omega r)^{2}}
  =
  \frac{\Omega r\,\delta u-U_{\mathrm{hub}}\,\delta v_t}
       {U_{\mathrm{hub}}^{2}+(\Omega r)^{2}},
\end{equation}
where $\delta v_t$ is the in-plane velocity perturbation along the tangential
direction of the blade path.
The transverse term therefore carries the factor $U_{\mathrm{hub}}/(\Omega r)$, which
is small over most of the span, whereas the longitudinal term is not scaled down.
Every point of the rotor disk is then a sample of the same scalar process $u$, and the
excitation follows from the cross-spectrum of $u$ between pairs of such points.


\subsection{Common IEC Kaimal Turbulence Basis}\label{sec:kaimal}


The IEC Kaimal model treats the fluctuating wind velocity as a stationary Gaussian
process whose spatial correlation is isotropic in three dimensions.
Based on the IEC Kaimal spectrum~\cite{IEC-2019}, all turbulence parameters are
referenced to $z_{\mathrm{hub}}$ and $U_{\mathrm{hub}}$, and the standard deviation
(STD) of this zero-mean spectrum is defined as:
\begin{equation}\label{eq:sigma1}
  \sigma_1=
  \begin{cases}
  I_{\mathrm{ref}}\, ( 0.75U_{\mathrm{hub}}+5.6), & \text{NTM}, \\[2pt]
  2I_{\mathrm{ref}}\, \left[0.072\left(\dfrac{V_{\mathrm{ave}}}{2}+3\right)
  \left(\dfrac{U_{\mathrm{hub}}}{2}-4\right)+10\right], & \text{ETM},\\[2pt]
  0.11U_{\mathrm{hub}}, & \text{EWM},
  \end{cases}
\end{equation}
where $V_{\mathrm{ave}}=0.2V_{\mathrm{ref}}$. The IEC scale parameter is
\begin{equation}\label{eq:Lambda1}
  \Lambda_1=
  \begin{cases}
  0.7z_{\mathrm{hub}}, & z_{\mathrm{hub}}\le 60\ \mathrm{m},\\
  42\ \mathrm{m}, & z_{\mathrm{hub}}>60\ \mathrm{m},
  \end{cases}
\end{equation}
and
\begin{equation}\label{eq:length-scales}
  \ell_u=8.1\Lambda_1,
  \qquad \ell_v=2.7\Lambda_1,
  \qquad \ell_w=0.66\Lambda_1,
\end{equation}
\begin{equation}\label{eq:sigma-components}
  \sigma_u=\sigma_1,
  \qquad \sigma_v=0.8\sigma_1,
  \qquad \sigma_w=0.5\sigma_1.
\end{equation}
\begin{equation}\label{eq:kaimal}
  S_k(f)=\frac{4\sigma_k^2\ell_k/U_{\mathrm{hub}}}
  {\left(1+6f\ell_k/U_{\mathrm{hub}}\right)^{5/3}},
  \qquad k\in\{u,v,w\}.
\end{equation}
The IEC Kaimal spectrum is specified as a one-sided spectrum. For concise
derivations, it is represented below by an even two-sided spectrum whose integral
over the whole frequency axis equals the longitudinal turbulence variance:
\begin{equation}\label{eq:kaimal-two-sided}
  S_u(f)
  =\frac{2\sigma_u^2\ell_u/U_{\mathrm{hub}}}
  {\left(1+6\lvert f\rvert\ell_u/U_{\mathrm{hub}}\right)^{5/3}},
  \quad f\in(-\infty,\infty),
  \qquad \int_{-\infty}^{\infty}S_u(f)\,\mathrm{d}f=\sigma_u^2.
\end{equation}
The one-sided form of the reported spectra is given in \cref{sec:numerical}.

All subsequent derivations of the rotor full CSD are built on the longitudinal
turbulence component $u$ alone,
the remaining components $v$ and $w$ are neglected
by default.
For any spatial distance $d$, IECKaimal spectral coherence gives
\begin{equation}\label{eq:iec-coherence}
  \gamma(f,d)=
  \exp\left[
  -12\sqrt{
  \left(\frac{fd}{U_{\mathrm{hub}}}\right)^2+
  \left(\frac{0.12d}{L_c}\right)^2}
  \right],
  \qquad L_c=8.1\Lambda_1.
\end{equation}
Based on \emph{Taylor's frozen-turbulence hypothesis},
the turbulent field is convected past the rotor
as a rigid pattern at $U_{\mathrm{hub}}$. The field is homogeneous, so every point shares the single-point
Kaimal spectrum $S_u(f)$ of \cref{eq:kaimal} and the IEC coherence
\cref{eq:iec-coherence} gives the spatial correlation.
The cross-spectrum is therefore real, which gives
\begin{equation}\label{eq:rotor-plane-csd}
  S_{ij}^{(u)}(f)=\gamma(f,d_{ij})\,S_u(f).
\end{equation}

It is worth noting that a real wind rotor is not strictly confined to a fixed plane:
yaw and nacelle tilt introduce an angular misalignment, and the blade geometry itself
is not strictly planar. These effects are neglected here; the yaw angle is ignored and
the rotor is assumed to span a strict plane, which leaves some physical discrepancy in
the present formulation.
Only the aerodynamic side of this approximation is relaxed downstream: the numerical
implementation of \cref{sec:numerical} accepts transfer coefficients that depend on
rotor azimuth in general, so the azimuthal load modulation that yaw, tilt, or blade
non-planarity produce enters the assembled CSD, while the coherence retains
the planar separation form of the rotor disk as illustrated in \cref{eq:separation}.


\subsection{Fixed-Frame Rotor-Plane Turbulence}\label{sec:fixed-frame}


For any two points at radial locations $r_i$ and $r_j$ in the rotor disk
separated by an angle $\theta$, their separation is
\begin{equation}\label{eq:separation}
  d_{ij}(\theta)=\sqrt{r_i^2+r_j^2-2r_i r_j\cos\theta}.
\end{equation}
Equation \cref{eq:separation} has two properties that follow directly from the
planar geometry of the rotor disk. It is \emph{periodic}: advancing the relative
angle by one full revolution returns the two points to the same relative position,
so the separation repeats itself,
\begin{equation}\label{eq:separation-periodic}
  d_{ij}(\theta+2\pi)=d_{ij}(\theta).
\end{equation}
It is also \emph{even}: since $\cos(-\theta)=\cos\theta$, a point $i$ leading
point $j$ by the angle $\theta$ and a point $i$ lagging it by $\theta$ give the
same separation,
\begin{equation}\label{eq:separation-even}
  d_{ij}(-\theta)=d_{ij}(\theta).
\end{equation}
Both properties are inherited by the coherence, which enters
the rotor-plane cross-spectrum only through $d_{ij}(\theta)$.
Consequently the map $\theta\mapsto\gamma\!\left(f,d_{ij}(\theta)\right)$ is a smooth
$2\pi$-periodic function of $\theta$ and therefore admits a uniformly convergent
Fourier series in the orthogonal complex-exponential basis
$\{\ee^{\mathrm{i}n\theta}\}_{n\in\mathbb{Z}}$,
\begin{equation}\label{eq:coherence-expansion}
  \gamma\left(f,d_{ij}(\theta)\right)
  =\sum_{n=-\infty}^{\infty}F_n(f;r_i,r_j)e^{\mathrm{i}n\theta},
\end{equation}
whose Fourier coefficients are defined by
\begin{equation}\label{eq:Fn}
 F_n(f;r_i,r_j)=
  \frac{1}{2\pi}\int_0^{2\pi}
  \gamma\left(f,d_{ij}(\theta)\right)e^{-\mathrm{i}n\theta} \, \mathrm{d} \theta
  =
  \frac{1}{\pi}\int_0^\pi
  \gamma\left(f,d_{ij}(\theta)\right)\cos(n\theta) \, \mathrm{d} \theta.
\end{equation}
%
%

The expansion is not a mere change of representation: as detailed in
\cref{sec:rot-sampling}, the angular order $n$ becomes a temporal frequency as the
rotor sweeps the disk.
Consequently, the coefficients $F_n(f; r_i, r_j)$ are the primary quantities
accumulated over radial stations,
blades, and harmonics in the subsequent derivations.



\subsection{Rotational Sampling and the Frequency-Shifted Kernel}\label{sec:rot-sampling}


Aerodynamic excitation arises from the frozen turbulence field sampled along
the helical trajectory of each blade element.
The core quantity required for the spectral formulation is the two-point 
cross-covariance between perturbations at station $(i,b)$ at time $t$
and station $(j,c)$ at time $t+\tau$:
\begin{equation}\label{eq:R_def}
  R_{ij}^{bc}(\tau) := \mathbb{E}\!\left[u_{ib}(t)\, u_{jc}^{\ast}(t+\tau)\right].
\end{equation}
Following the generalized Fourier transform convention for cross-spectral density,
\begin{equation}\label{eq:covariance-convention}
  R_{xy}(\tau) 
  = 
  \int_{-\infty}^{\infty} S_{xy}(f)e^{\mathrm{i}2\pi f\tau}\,\mathrm{d}f .
\end{equation}

Under the assumption of homogeneous turbulence,
the cross-spectrum in the fixed frame is 
completely specified by the coherence expansion \cref{eq:coherence-expansion}. 
Blade rotation affects the statistics solely through
the time-lag-dependent spatial separation of the two elements. 
Let the azimuth of blade $b$ ($b = 1, \dots, B$) at time $t$ be
\begin{equation}\label{eq:azimuth}
  \psi_b(t) = \psi_0 + \beta_b + \Omega t, 
  \qquad 
  \beta_b = \frac{2\pi(b-1)}{B},
\end{equation}
where $\beta_b$ is the initial blade azimuthal angle. 
The signed angular separation between element $(i,b)$ at time $t$
and element $(j,c)$ at time $t+\tau$ is then
\begin{equation}\label{eq:theta-bc}
  \theta_{bc}(\tau) 
  := \psi_c(t+\tau) - \psi_b(t) 
  = \Delta\beta_{bc} + \Omega\tau,
\end{equation}
with $\Delta\beta_{bc} := \beta_c - \beta_b$.

For a fixed turbulence frequency $\nu$ in the fixed frame,
the axisymmetric coherence depends on the rotor-disk 
geometry solely through the angular separation $\theta$.
The Fourier series \cref{eq:coherence-expansion} decomposes
this angular dependence into spatial harmonics of the form 
$e^{\mathrm{i}n\theta}$ ($n \in \mathbb{Z}$).

Rotor rotation converts this spatial variation into a temporal modulation. 
Substituting \cref{eq:theta-bc} into the $n$th coherence harmonic yields
\begin{equation}\label{eq:harmonic-contribution}
  F_n(\nu; r_i, r_j) e^{\mathrm{i}n\theta_{bc}(\tau)}
  =
  \underbrace{F_n(\nu; r_i, r_j) e^{\mathrm{i}n\Delta\beta_{bc}}}_{\text{lag-independent phase shift}}
  \cdot
  \underbrace{e^{\mathrm{i} 2\pi n f_{\mathrm{rot}}\tau}}_{\text{harmonic frequency shift}},
\end{equation}
where $\Omega = 2\pi f_{\mathrm{rot}}$,
with $f_{\mathrm{rot}}$ denoting
the fundamental rotational frequency of the rotor. 
Equation \cref{eq:harmonic-contribution} explicitly decouples
the inter-blade geometric phase difference 
$n\Delta\beta_{bc}$ from the temporal harmonic modulation
$e^{\mathrm{i}2\pi n f_{\mathrm{rot}}\tau}$.
To distinguish the fixed-frame turbulence frequency from the rotating-frame observation frequency $f$,
denote the fixed-frame frequency by $\nu$.
Substituting \cref{eq:coherence-expansion} and
\cref{eq:harmonic-contribution} into the cross-covariance
definition \cref{eq:covariance-convention}
then expresses $R_{ij}^{bc}(\tau)$ in terms of the coherence
harmonics $F_n$:
\begin{equation}\label{eq:Rij-harmonic-expansion}
  R_{ij}^{bc}(\tau)
  =
  \sum_{n=-\infty}^{\infty}
  e^{\mathrm{i}n\Delta\beta_{bc}}
  \int_{-\infty}^{\infty}
  F_n(\nu; r_i, r_j) S_u(\nu)\,
  e^{\mathrm{i} 2\pi (\nu + n f_{\mathrm{rot}})\tau}
  \,\mathrm{d}\nu .
\end{equation}
The exponent in \cref{eq:Rij-harmonic-expansion} reflects a purely kinematic
modulation: the coherence seen by a blade-element pair is $2\pi$-periodic in the
separation angle $\theta_{bc}$, so its $n$th harmonic completes $n$ cycles per
revolution and the observed frequency is displaced by integer multiples of
$f_{\mathrm{rot}}$~\cite{Dragt-1984,Dragt-1985,Halfpenny-1998-FD}:
\begin{equation}\label{eq:sampled-wind-frequency}
  f = \nu + n f_{\mathrm{rot}}
  \quad \Longleftrightarrow \quad
  \nu = f - n f_{\mathrm{rot}}.
\end{equation}
The apparent frequency observed in the rotating frame is therefore $f$,
which requires the fixed-frame turbulence spectrum and coherence to be
evaluated at the Doppler-shifted argument $\nu = f - n f_{\mathrm{rot}}$.
To explicitly separate the spatial coherence harmonic from the temporal
frequency shift, Halfpenny~\cite{Halfpenny-1998-FD} formulated the
two-index \emph{Rotational-Sampling Kernel}:
\begin{equation}\label{eq:kernel}
  K_{ij}^{(n,q)}(f)
  :=
  F_n\!\left(f - q f_{\mathrm{rot}}; r_i, r_j\right)
  S_u\!\left(f - q f_{\mathrm{rot}}\right),
\end{equation}
where the two indices $n$ and $q$ denote distinct physical mechanisms:
\begin{enumerate}
  \item The \emph{spatial harmonic index} $n \in \mathbb{Z}$ isolates the $n$th circumferential wavenumber from the azimuthal Fourier expansion of spatial coherence \cref{eq:coherence-expansion}.
  \item The \emph{temporal modulation order} $q \in \mathbb{Z}$ governs the Doppler frequency shift, mapping the observed frequency $f$ to the fixed-frame frequency $\nu = f - q f_{\mathrm{rot}}$.
\end{enumerate}

Whether these two indices coincide is determined by the lag-dependent phase of the
cross-covariance.
For pure wind sampling, the integrand of \cref{eq:Rij-harmonic-expansion} exhibits the
lag phase $e^{\mathrm{i}2\pi(\nu + n f_{\mathrm{rot}})\tau}$.
Matching this carrier with the spectral definition $e^{\mathrm{i}2\pi f\tau}$ defines
the observation frequency $f = \nu + n f_{\mathrm{rot}}$.
Comparing this result directly with the kernel's frequency-shift relation
$f = \nu + q f_{\mathrm{rot}}$ in \cref{eq:kernel} yields the exact identity
\begin{equation}\label{eq:q-equals-n-proof}
  \underbrace{q \equiv n}_{\text{pure wind sampling}} .
\end{equation}
This equality holds because the only source of time lag $\tau$ is the
rigid kinematic
trajectory $\theta_{bc}(\tau) = \Delta\beta_{bc} + \Omega\tau$
in the argument of spatial
coherence; no additional time-dependent multiplier
acts on the inflow perturbation $u$.
Consequently, for undisturbed wind sampling,
the two-index kernel collapses onto its
diagonal:
\begin{equation}\label{eq:kernel-sampled-wind}
  K_{ij}^{(n,n)}(f)
  =
  F_n\!\left(f - n f_{\mathrm{rot}}; r_i, r_j\right)
  S_u\!\left(f - n f_{\mathrm{rot}}\right).
\end{equation}

In contrast, the condition $q \neq n$
arises when an azimuth-dependent multiplier modulates
the sampled signal.
As developed in \cref{sec:six-component-sensitivity},
mapping local inflow to aerodynamic loads
introduces transfer coefficients with Fourier harmonics $e^{\mathrm{i}s\psi}$.
The product of such coefficients evaluated at $t$ and $t+\tau$
contributes an additional
lag factor $e^{-\mathrm{i}2\pi s f_{\mathrm{rot}}\tau}$,
shifting the net phase carrier to
$e^{\mathrm{i}2\pi(n-s)f_{\mathrm{rot}}\tau}$ and establishing
\begin{equation}\label{eq:q-equals-n-minus-s-preview}
  \underbrace{q \equiv n - s}_{\text{azimuthal load modulation}} .
\end{equation}
Whenever $s \neq 0$ (such as for transverse hub loads where $s = \pm 1$, or under
aerodynamic asymmetries), the Doppler shift order $q$ decouples from the spatial coherence
order $n$.
This distinction necessitates retaining the general two-index notation $K_{ij}^{(n,q)}(f)$
in \cref{eq:kernel}, even though wind sampling utilizes only the diagonal $q = n$.



\subsection{Aerodynamic Load-Transfer Coefficients}\label{sec:six-component-sensitivity}


With the inflow statistics specified, the hub loads follow from their
aerodynamic transmission into the non-rotating hub-fixed frame.
Following Halfpenny~\cite{Halfpenny-1998-FD},
this mapping is formulated by linearizing the aerodynamic loads
about the steady operating state.
For blade $b$ at azimuth $\psi$, the spanwise inflow across the $N_r$ radial
stations
is decomposed into a steady operating inflow vector $\overline{\boldsymbol{u}}$
and a zero-mean perturbation vector $\delta\boldsymbol{u}(\psi)$:
\begin{equation}\label{eq:local-inflow-vector}
  \boldsymbol{u}(\psi)
  =
  \overline{\boldsymbol{u}}
  +
  \delta\boldsymbol{u}(\psi),
  \qquad
  \delta\boldsymbol{u}
  =
  \begin{bmatrix}
    \delta u_1 & \cdots & \delta u_{N_r}
  \end{bmatrix}^{T},
\end{equation}
where $\delta u_i$ denotes the longitudinal inflow perturbation
at radial station $i$ ($i=1,\ldots,N_r$).

The six hub-load components are collected in the hub-fixed load vector:
\begin{equation}\label{eq:hub-load-vector-def}
  \boldsymbol{L} =
  \begin{bmatrix} F_x & F_y & F_z & Q_x & Q_y & Q_z \end{bmatrix}^{\mathrm{T}},
\end{equation}
where $F_x$ is the axial thrust, $Q_x$ is the shaft torque, $F_y$ and $F_z$ are the
in-plane lateral and vertical shear forces, and $Q_y$ and $Q_z$ are the hub nodding
(pitching) and yawing moments, respectively. The component index set $\mathcal{L}$
consists of the entries of $\boldsymbol{L}$.

The perturbation of load component $a\in\mathcal{L}$
contributed by radial station $i$
on blade $b$ is expressed as
\begin{equation}\label{eq:delta-F}
  \delta F^{(a)}_{ib}(t)
  =
  g_i^{(a)}(\psi_b(t))\,\delta u_{ib}(t),
\end{equation}
where $g_i^{(a)}(\psi)$ is the azimuth-dependent load-transfer coefficient
(or aerodynamic gain factor defined by Halfpenny~\cite{Halfpenny-1998-FD})
mapping local inflow fluctuations into the hub-fixed frame.
Superposing \cref{eq:delta-F} over all $N_r$ radial stations
and all $B$ blades yields the total hub-load perturbation:
\begin{equation}\label{eq:total-rotor-load}
  \delta F^{(a)}(t)
  =
  \sum_{b=1}^{B}\sum_{i=1}^{N_r}
  \delta F^{(a)}_{ib}(t)
  =
  \sum_{b=1}^{B}\sum_{i=1}^{N_r}
  g_i^{(a)}(\psi_b(t))\,\delta u_{ib}(t).
\end{equation}
Under steady rotor rotation at angular speed
$\Omega = 2\pi f_{\mathrm{rot}}$,
the azimuth of blade $b$ follows \cref{eq:azimuth}.
Since the kinematic transformation to the non-rotating hub frame
and the steady operating state are time-invariant,
the resulting sensitivity operator is strictly
$2\pi$-periodic with respect to rotor azimuth:
\begin{equation}\label{eq:sensitivity-periodicity}
  g_i^{(a)}(\psi+2\pi)
  =
  g_i^{(a)}(\psi).
\end{equation}
At the steady operating point with fixed rotor speed,
$g_i^{(a)}(\psi_b(t))$ is deterministic and periodic,
so the product $g_i^{(a)}(\psi_b(t))\,\delta u_{ib}(t)$ is cyclostationary.
To express this modulation in the frequency domain and combine it with
the spatial coherence harmonics of \cref{sec:rot-sampling},
$g_i^{(a)}(\psi)$ is expanded into a complex azimuthal Fourier series:
\begin{equation}\label{eq:sensitivity-expansion}
  g_i^{(a)}(\psi)
  =
  \sum_{s=-\infty}^{\infty}
  \hat{g}_{i,s}^{(a)}e^{\mathrm{i}s\psi},
  \qquad
  \hat{g}_{i,s}^{(a)}
  =
  \frac{1}{2\pi}
  \int_0^{2\pi}
  g_i^{(a)}(\psi)
  e^{-\mathrm{i}s\psi}
  \,\mathrm{d}\psi,
\end{equation}
where $s\in\mathbb{Z}$ represents the azimuthal harmonic order of the load-transfer
mechanism. Along the trajectory of blade $b$, this expansion yields
$g_i^{(a)}(\psi_b(t)) = \sum_s \hat{g}_{i,s}^{(a)} e^{\mathrm{i}s(\psi_0+\beta_b)} e^{\mathrm{i}2\pi s f_{\mathrm{rot}}t}$,
which explicitly decouples the initial rotor orientation $\psi_0$, the blade spacing $\beta_b$,
and the temporal modulation frequency $s f_{\mathrm{rot}}$.


\subsection{Derivation of Aerodynamic Load-Transfer Coefficients
Based on Blade Element Momentum Theory}\label{sec:bemt-sensitivities}


To provide an explicit physical foundation for the load-transfer coefficients
$g_i^{(a)}(\psi)$ introduced in \cref{sec:six-component-sensitivity},
the sectional aerodynamic forces are resolved onto the hub-fixed reference frame.
Consider element $i$ ($i = 1, \dots, N_r$) at radial position $r_i$
on blade $b$ ($b = 1, \dots, B$).
The local aerodynamic resultant is characterized by
the sectional normal force $N_{ib}$
(aligned with the rotor shaft axis $+\boldsymbol{e}_x$)
and the sectional in-plane tangential force $T_{ib}$
(acting in the direction of increasing rotor azimuth $\psi_b$).
The sectional aerodynamic pitching moment $M_{ib}$ is neglected,
as its integrated contribution to the hub moment is negligible
compared to the moment arms of $N_{ib}$ and $T_{ib}$.

Based on the blade-element momentum method dervied by Andrew Ning et al.
\cite{CCBlade1,CCBlade2},
local forces are driven by
the longitudinal inflow perturbation $\delta u_{ib}(t)$.
Linearization about the operating mean state gives
\begin{equation}\label{eq:sectional-linearization}
  \delta N_{ib}(t) = h_{N,i}(\psi_b)\,\delta u_{ib}(t),
  \qquad
  \delta T_{ib}(t) = h_{T,i}(\psi_b)\,\delta u_{ib}(t),
\end{equation}
where the sectional sensitivities
$h_{N,i} \coloneqq \partial N_i / \partial u_i$ and
$h_{T,i} \coloneqq \partial T_i / \partial u_i$
represent the aerodynamic derivatives evaluated with respect to
the local inflow perturbations while
keeping all other operating parameters fixed.

Resolving these element forces onto the hub-fixed non-rotating axes at
instantaneous azimuth $\psi_b$ gives the element contribution vector
$\boldsymbol{\ell}_i(\psi_b)$.
Summing these contributions gives the hub-load vector defined in
\cref{eq:hub-load-vector-def}:
\begin{equation}\label{eq:element-load-contribution}
  \boldsymbol{\ell}_i(\psi_b) \coloneqq
  \begin{bmatrix}
    N_i(\psi_b) \\
    -T_i(\psi_b)\sin\psi_b \\
    T_i(\psi_b)\cos\psi_b \\
    T_i(\psi_b)r_i \\
    N_i(\psi_b)r_i\sin\psi_b \\
    -N_i(\psi_b)r_i\cos\psi_b
  \end{bmatrix},
  \qquad
  \boldsymbol{L}(t) = \sum_{b=1}^{B}\sum_{i=1}^{N_r}\boldsymbol{\ell}_i(\psi_b(t)).
\end{equation}
Expressing the hub-load perturbation caused by element $(i, b)$
as $\delta F^{(a)}_{ib}(t) = g_i^{(a)}(\psi_b(t))\,\delta u_{ib}(t)$
defines the six-component single-blade transfer sensitivities:
\begin{equation}\label{eq:sensitivities-def}
  \begin{aligned}
    g_i^{(F_x)}(\psi) &= h_{N,i}(\psi), 
    & g_i^{(Q_x)}(\psi) &= h_{T,i}(\psi)r_i, \\
    g_i^{(F_y)}(\psi) &= -h_{T,i}(\psi)\sin\psi, 
    & g_i^{(Q_y)}(\psi) &= h_{N,i}(\psi)r_i\sin\psi, \\
    g_i^{(F_z)}(\psi) &= h_{T,i}(\psi)\cos\psi, 
    & g_i^{(Q_z)}(\psi) &= -h_{N,i}(\psi)r_i\cos\psi.
  \end{aligned}
\end{equation}

When the base sectional aerodynamic derivatives $h_{N,i}$ and $h_{T,i}$ are azimuth-independent (e.g., under uniform axial inflow without steady shear or tilt), the directional projections $\cos\psi$ and $\sin\psi$ restrict the complex Fourier series $g_i^{(a)}(\psi) = \sum_{s} \hat{g}_{i,s}^{(a)} e^{\mathrm{i} s \psi}$ in \cref{eq:sensitivity-expansion} to discrete harmonic orders $s \in \{0, \pm 1\}$:
\begin{equation}\label{eq:fourier-coefficients-explicit}
  \begin{aligned}
    \hat{g}_{i,0}^{(F_x)} &= h_{N,i}, 
    & \hat{g}_{i,0}^{(Q_x)} &= h_{T,i}r_i, \\
    \hat{g}_{i,+1}^{(F_y)} &= \frac{\mathrm{i}}{2}h_{T,i}, 
    & \hat{g}_{i,-1}^{(F_y)} &= -\frac{\mathrm{i}}{2}h_{T,i}, \\
    \hat{g}_{i,+1}^{(F_z)} &= \frac{1}{2}h_{T,i}, 
    & \hat{g}_{i,-1}^{(F_z)} &= \frac{1}{2}h_{T,i}, \\
    \hat{g}_{i,+1}^{(Q_y)} &= -\frac{\mathrm{i}}{2}h_{N,i}r_i, 
    & \hat{g}_{i,-1}^{(Q_y)} &= \frac{\mathrm{i}}{2}h_{N,i}r_i, \\
    \hat{g}_{i,+1}^{(Q_z)} &= -\frac{1}{2}h_{N,i}r_i, 
    & \hat{g}_{i,-1}^{(Q_z)} &= -\frac{1}{2}h_{N,i}r_i.
  \end{aligned}
\end{equation}


\subsection{Derivation of Hub-Load Cross-Covariance}\label{sec:load-covariance-derivation}


\subsubsection{Index Reduction via Azimuthal Averaging and Blade Symmetry}\label{sec:load-harmonic-order}

To evaluate the spectral density of the total rotor load, the second-order
cross-covariance of \cref{eq:total-rotor-load} is formulated. Because the
load-transfer coefficients are deterministic at the operating state, the
statistical expectation acts solely on the turbulent inflow field, giving
\begin{align}\label{eq:R-ad}
  R_{ad}(t,\tau)
  &=
  \mathbb{E}\!\left[
    \delta F^{(a)}(t)\,\delta F^{(d)\ast}(t+\tau)
  \right]
  \nonumber\\
  &=
  \sum_{b,c=1}^B\sum_{i,j=1}^{N_r}
  \mathbb{E}\!\left[
    g_i^{(a)}(\psi_b(t))\delta u_{ib}(t)\,
    g_j^{(d)\ast}(\psi_c(t+\tau))\delta u_{jc}^{\ast}(t+\tau)
  \right]
  \nonumber\\
  &=
  \sum_{b,c=1}^B\sum_{i,j=1}^{N_r}
  g_i^{(a)}(\psi_b(t))\,
  g_j^{(d)\ast}(\psi_c(t+\tau))
  R_{ij}^{bc}(\tau),
\end{align}
where $R_{ij}^{bc}(\tau) = \mathbb{E}[\delta u_{ib}(t)\delta u_{jc}^\ast(t+\tau)]$
is the blade-resolved wind cross-covariance from \cref{eq:R_def}.
Expanding the two sensitivity coefficients at their respective time endpoints
via \cref{eq:sensitivity-expansion} yields
\begin{equation}\label{eq:sensitivity-fourier-pair}
  g_i^{(a)}(\psi_b(t))
  =
  \sum_{s=-\infty}^{\infty} \hat{g}_{i,s}^{(a)} e^{\mathrm{i}s\psi_b(t)},
  \qquad
  g_j^{(d)\ast}(\psi_c(t+\tau))
  =
  \sum_{r=-\infty}^{\infty} \hat{g}_{j,r}^{(d)\ast} e^{-\mathrm{i}r\psi_c(t+\tau)},
\end{equation}
where the azimuthal harmonic orders $s, r \in \mathbb{Z}$ are initially independent.
With the lagged blade azimuth $\psi_c(t+\tau) = \psi_b(t) + \Delta\beta_{bc} + \Omega\tau$
from \cref{eq:theta-bc}, the combined aerodynamic phase evaluates to
\begin{align}\label{eq:azimuthal-harmonic-product}
  e^{\mathrm{i}s\psi_b(t)}
  e^{-\mathrm{i}r\psi_c(t+\tau)}
  &=
  e^{\mathrm{i}(s-r)\psi_b(t)}
  e^{-\mathrm{i}r\Delta\beta_{bc}}
  e^{-\mathrm{i}r\Omega\tau}
  \nonumber\\
  &=
  e^{\mathrm{i}(s-r)\psi_b(t)}
  e^{-\mathrm{i}r\Delta\beta_{bc}}
  e^{-\mathrm{i}2\pi r f_{\mathrm{rot}}\tau}.
\end{align}
Multiplying \cref{eq:azimuthal-harmonic-product} by the spatial coherence phase
$e^{\mathrm{i}n\Delta\beta_{bc}}e^{\mathrm{i}2\pi n f_{\mathrm{rot}}\tau}$ from
the wind field expansion \cref{eq:Rij-harmonic-expansion} yields the complete
phase carrier for a generic order triplet $(n,s,r)$:
\begin{align}\label{eq:generic-phase}
  &e^{\mathrm{i}s\psi_b(t)}
  e^{-\mathrm{i}r\psi_c(t+\tau)}
  e^{\mathrm{i}n\Delta\beta_{bc}}
  e^{\mathrm{i}n\Omega\tau}
  \nonumber\\
  &\qquad=
  e^{\mathrm{i}(s-r)\psi_b(t)}
  e^{\mathrm{i}(n-r)\Delta\beta_{bc}}
  e^{\mathrm{i}2\pi(n-r)f_{\mathrm{rot}}\tau}.
\end{align}
Substituting \cref{eq:generic-phase} into the cross-covariance leads to the full
harmonic representation:
\begin{equation}\label{eq:generic-covariance}
  \begin{aligned}
  R_{ad}(t,\tau)
  ={} & \sum_{b,c=1}^{B}\sum_{i,j=1}^{N_r}\sum_{n,s,r=-\infty}^{\infty}
        \hat{g}_{i,s}^{(a)}\hat{g}_{j,r}^{(d)\ast}
        e^{\mathrm{i}(s-r)\psi_b(t)}e^{\mathrm{i}(n-r)\Delta\beta_{bc}} \\
      & \times \int_{-\infty}^{\infty}F_n(\nu;r_i,r_j)\,S_u(\nu)\,
        e^{\mathrm{i}2\pi[\nu+(n-r)f_{\mathrm{rot}}]\tau}\,\mathrm{d}\nu .
  \end{aligned}
\end{equation}
Because $\psi_b(t) = \psi_0 + \beta_b + \Omega t$, the presence of $e^{\mathrm{i}(s-r)\psi_b(t)}$
renders $R_{ad}(t,\tau)$ cyclostationary in $t$ and dependent on the reference azimuth $\psi_0$.
The stationary (zeroth-cyclic) covariance is extracted by averaging over $\psi_0$:
\begin{equation}\label{eq:stationary-covariance}
  \overline{R}_{ad}(\tau)
  =
  \frac{1}{2\pi}
  \int_0^{2\pi}
  R_{ad}(t,\tau)\,\mathrm{d}\psi_0.
\end{equation}
Fourier orthogonality over the initial orientation,
\begin{equation}\label{eq:fourier-orthogonality}
  \frac{1}{2\pi}
  \int_0^{2\pi}
  e^{\mathrm{i}(s-r)\psi_0}\,\mathrm{d}\psi_0
  =
  \delta_{sr},
\end{equation}
strictly eliminates all cross-harmonic terms $s \neq r$, retaining only the matched
harmonics $r = s$, for which the phase factor $e^{\mathrm{i}(s-r)\psi_b(t)}$ is unity.
\Cref{eq:generic-covariance} then collapses to the strictly stationary form:
\begin{equation}\label{eq:stationary-R-ad-collapsed}
  \begin{aligned}
  \overline{R}_{ad}(\tau)
  ={} & \sum_{b,c=1}^{B}\sum_{i,j=1}^{N_r}\sum_{n,s=-\infty}^{\infty}
        \hat{g}_{i,s}^{(a)}\hat{g}_{j,s}^{(d)\ast}
        e^{\mathrm{i}(n-s)\Delta\beta_{bc}} \\
      & \times \int_{-\infty}^{\infty}
        \underbrace{F_n(\nu;r_i,r_j)\,S_u(\nu)}_{K_{ij}^{(n,q)}}
        e^{\mathrm{i}2\pi[\nu+(n-s)f_{\mathrm{rot}}]\tau}\,\mathrm{d}\nu .
  \end{aligned}
\end{equation}
Matching the resulting lag phase $e^{\mathrm{i}2\pi[\nu+(n-s)f_{\mathrm{rot}}]\tau}$ with
the observed output frequency modulation $f = \nu + q f_{\mathrm{rot}}$ directly establishes
the exact identity
\begin{equation}\label{eq:q-equals-n-minus-s-again}
  q \equiv n - s.
\end{equation}
Consequently, the spatial coherence harmonic $n$ and the aerodynamic sensitivity harmonic $s$
superimpose to dictate the net temporal frequency shift $q$ in the fixed hub frame, evaluated at
\begin{equation}\label{eq:shifted-field-frequency}
  f - q f_{\mathrm{rot}}
  =
  f - (n - s)f_{\mathrm{rot}}.
\end{equation}
Next, rotor rotational symmetry acts on the blade-pair phase factor
$e^{\mathrm{i}(n-s)\Delta\beta_{bc}} = e^{\mathrm{i}q\Delta\beta_{bc}}$.
For $B$ identical, equally spaced blades, this summation decouples into the modulus squared
of a geometric series~\cite{Halfpenny-1998-FD}:
\begin{align}\label{eq:blade-pair-sum-closed}
  \sum_{b=1}^{B}\sum_{c=1}^{B} e^{\mathrm{i}q\Delta\beta_{bc}}
  &=
  \left(\, \sum_{c=1}^{B}e^{\mathrm{i}q\beta_c} \,\right)
  \left(\,\sum_{b=1}^{B}e^{-\mathrm{i}q\beta_b} \,\right)
  \nonumber\\
  &=
  \left|\sum_{b=1}^{B} e^{\mathrm{i}q\beta_b}\right|^2
  =
  \left|\sum_{m=0}^{B-1} e^{\mathrm{i}2\pi qm/B}\right|^2
  \nonumber\\
  &=
  \begin{cases}
    B^2, & q = pB, \\
    0,   & q \neq pB,
  \end{cases}
  \qquad p \in \mathbb{Z},
\end{align}
where $p$ denotes the blade-passing harmonic index. Rotor symmetry therefore enforces $q = pB$.
Combining this with $q = n - s$ yields the dual harmonic selection rules:
\begin{equation}\label{eq:q-and-n-boxed}
  q = pB,
  \qquad
  n = pB + s,
  \qquad
  \nu = f - pBf_{\mathrm{rot}}.
\end{equation}
The distinction is fundamental: blade symmetry selects the temporal frequency order $q = pB$
in the hub frame, whereas the spatial coherence order $n = pB + s$ is shifted by the aerodynamic
harmonic order $s$. The rotational-sampling kernel is therefore identified as
\begin{equation}\label{eq:kernel-index-equivalence}
  K_{ij}^{(n,q)}(f)
  \equiv
  K_{ij}^{(pB+s,\, pB)}(f).
\end{equation}
Substituting \cref{eq:blade-pair-sum-closed} and \cref{eq:q-and-n-boxed} into
\cref{eq:stationary-R-ad-collapsed} yields the blade-summed stationary cross-covariance:
\begin{equation}\label{eq:stationary-covariance-blade-summed}
  \overline{R}_{ad}(\tau)
  =
  B^2
  \sum_{p=-\infty}^{\infty}
  \sum_{i,j=1}^{N_r}
  \sum_{s=-\infty}^{\infty}
  \hat{g}_{i,s}^{(a)}
  \hat{g}_{j,s}^{(d)\ast}
  \int_{-\infty}^{\infty}
  K_{ij}^{(pB+s,pB)}(f)\,
  e^{\mathrm{i}2\pi f\tau}\,\mathrm{d}f.
\end{equation}
%
%
\subsubsection{Secondary Modulation Mechanisms: Axial Invariance and Transverse Decoupling}~\label{sec:secondary-modulation}

Before the complete 6-DOF hub-load CSD is assembled,
the aerodynamic modulation order is separated from the rotor Doppler effect.
It enters as a secondary modulation that acts on the carrier already fixed by
rotor sampling, rather than setting the Doppler shift itself. The orders retained
by the axial and transverse hub loads differ, and the two cases are treated in
turn below.

\begin{enumerate}[leftmargin=*]
  \item \textbf{Axial thrust and torque} ($F_x, Q_x$): In an idealized axisymmetric inflow,
  the load transfer is azimuth-independent ($s = 0$), so the spatial and temporal orders coincide:
  \begin{equation}\label{eq:axial-orders}
    F_x, Q_x:
    \qquad
    s = 0,
    \qquad
    n = q = pB.
  \end{equation}

  \item \textbf{Transverse hub loads} ($F_y, F_z, Q_y, Q_z$): While the hub-load components
  indexed by $\mathcal{L}$ are strictly real Cartesian quantities, the
  kinematic transformation from rotating blade coordinates to the hub-fixed frame projects the
  transverse sensitivities $g_i^{(a)}(\psi)$ ($a \in \{F_y, F_z, Q_y, Q_z\}$) through
  directional trigonometric factors $\cos\psi$ and $\sin\psi$. Applying Euler's formula,
  \begin{equation}\label{eq:euler-projection}
    \cos\psi = \frac{e^{\mathrm{i}\psi} + e^{-\mathrm{i}\psi}}{2},
    \qquad
    \sin\psi = \frac{e^{\mathrm{i}\psi} - e^{-\mathrm{i}\psi}}{2\mathrm{i}},
  \end{equation}
  reveals that each individual real Cartesian transverse component contains a mixed conjugate
  harmonic pair $s = \pm 1$, as demonstrated explicitly in \cref{eq:fourier-coefficients-explicit}.
  Direct evaluation of the transverse cross-covariance in this real Cartesian basis produces
  cumbersome cross-product trigonometric terms that obscure the underlying rotational-sampling mechanisms.

  The complex circular load combinations separate these conjugate harmonics into
  single-harmonic exponential carriers:
  \begin{equation}\label{eq:circular-components}
    F_\pm \coloneqq F_y \pm \mathrm{i}F_z,
    \qquad
    Q_\pm \coloneqq Q_y \pm \mathrm{i}Q_z.
  \end{equation}
  Under the present azimuth, complex-plane, and time conventions, $F_+$ and $Q_+$ are the
  co-rotating components, whereas $F_-$ and $Q_-$ are the counter-rotating
  components in the rotor disk plane. Combining the transfer sensitivities
  \cref{eq:sensitivities-def} through \cref{eq:circular-components}, the conjugate
  exponentials of \cref{eq:euler-projection} cancel, and each circular transfer
  coefficient reduces to a single exponential term:
  \begin{equation}\label{eq:circular-sensitivity-decoupling}
    \begin{aligned}
      g_i^{(F_+)}(\psi) &= \mathrm{i}h_{T,i}\,e^{\mathrm{i}\psi},
      & g_i^{(F_-)}(\psi) &= -\mathrm{i}h_{T,i}\,e^{-\mathrm{i}\psi}, \\
      g_i^{(Q_+)}(\psi) &= -\mathrm{i}h_{N,i}r_i\,e^{\mathrm{i}\psi},
      & g_i^{(Q_-)}(\psi) &= \mathrm{i}h_{N,i}r_i\,e^{-\mathrm{i}\psi}.
    \end{aligned}
  \end{equation}
  Substituting the coefficients \cref{eq:fourier-coefficients-explicit} into the
  single-harmonic forms of \cref{eq:circular-sensitivity-decoupling} gives the
  Fourier coefficients of the circular transfer coefficients:
  \begin{equation}\label{eq:circular-azimuthal-orders}
    \begin{aligned}
      \hat{g}_{i,+1}^{(F_+)} &= \mathrm{i}h_{T,i},
      & \hat{g}_{i,+1}^{(F_-)} &= 0, \\
      \hat{g}_{i,-1}^{(F_+)} &= 0,
      & \hat{g}_{i,-1}^{(F_-)} &= -\mathrm{i}h_{T,i}, \\
      \hat{g}_{i,+1}^{(Q_+)} &= -\mathrm{i}h_{N,i}r_i,
      & \hat{g}_{i,+1}^{(Q_-)} &= 0, \\
      \hat{g}_{i,-1}^{(Q_+)} &= 0,
      & \hat{g}_{i,-1}^{(Q_-)} &= \mathrm{i}h_{N,i}r_i.
    \end{aligned}
  \end{equation}
  Combined with the blade comb filtering $q = pB$, the inflow spatial coherence orders
  $n = q + s = pB + s$ are uniquely determined:
  \begin{equation}\label{eq:transverse-orders}
    \begin{aligned}
      F_+, Q_+ \colon \quad &q = pB, \quad n = pB + 1, \\
      F_-, Q_- \colon \quad &q = pB, \quad n = pB - 1.
    \end{aligned}
  \end{equation}
  The corresponding two-index rotational-sampling kernels take the explicit form
  \begin{equation}\label{eq:transverse-kernels}
    \begin{aligned}
      K_{ij}^{(pB+1, \, pB)}(f)
      &=
      F_{pB+1}\!\left(f - pB f_{\mathrm{rot}}; r_i, r_j\right)
      S_u\!\left(f - pB f_{\mathrm{rot}}\right), \\
      K_{ij}^{(pB-1,\, pB)}(f)
      &=
      F_{pB-1}\!\left(f - pB f_{\mathrm{rot}}; r_i, r_j\right)
      S_u\!\left(f - pB f_{\mathrm{rot}}\right),
    \end{aligned}
  \end{equation}
  where $pB \pm 1$ designates the \emph{spatial coherence harmonic} entering the transverse load,
  while the temporal frequency shift remains centered at the blade-passing frequency $pB f_{\mathrm{rot}}$.

  Once the auto- and cross-spectral densities of the decoupled
  complex quantities $S_{F_+ F_+}(f)$,
  $S_{F_- F_-}(f)$, and $S_{F_+ F_-}(f)$
  are evaluated, 
  the physical real Cartesian hub load spectra
  are recovered through the elementary linear inverse mapping:
  \begin{equation}\label{eq:cartesian-recovery}
    \begin{aligned}
      F_y = \frac{F_+ + F_-}{2},
      &\qquad
      F_z = \frac{F_+ - F_-}{2\mathrm{i}}, \\[2pt]
      Q_y = \frac{Q_+ + Q_-}{2},
      &\qquad
      Q_z = \frac{Q_+ - Q_-}{2\mathrm{i}},
    \end{aligned}
  \end{equation}
  which avoids mixed trigonometric cross-terms during intermediate integrations. The
  same inverse mapping applies to the corresponding Cartesian spectra below.
\end{enumerate}


\subsection{Evaluation of the Rotor-Load CSD for an Ideal Rotor}\label{sec:hub-load-csd-matrix}


Applying the inverse Fourier transform identity of \cref{eq:csd-definition} to the
blade-summed stationary cross-covariance \cref{eq:stationary-covariance-blade-summed},
\begin{equation}
  \overline{R}_{ad}(\tau) = \int_{-\infty}^{\infty} S_{ad}(f) e^{\mathrm{i}2\pi f\tau}\,\mathrm{d}f,
\end{equation}
and identifying the carrier-shifted frequency $f = \nu + pB f_{\mathrm{rot}}$ directly yields the rotor-load cross-spectral density (CSD) for any $a, d \in \mathcal{L}$:
\begin{equation}\label{eq:rotor-load-csd}
  \begin{aligned}
    S_{ad}^{\mathrm{rotor}}(f)
    &=
    B^2
    \sum_{p=-\infty}^{\infty}
    \sum_{i,j=1}^{N_r}
    \sum_{s=-\infty}^{\infty}
    \hat{g}_{i,s}^{(a)}
    \hat{g}_{j,s}^{(d)\ast}
    F_{pB+s}\!\left(f - pB f_{\mathrm{rot}}; r_i, r_j\right)
    S_u\!\left(f - pB f_{\mathrm{rot}}\right)\\
    &=
    B^2
    \sum_{p=-\infty}^{\infty}
    \sum_{i,j=1}^{N_r}
    \sum_{s=-\infty}^{\infty}
    \hat{g}_{i,s}^{(a)}
    \hat{g}_{j,s}^{(d)\ast}
    K_{ij}^{(pB+s,\,pB)}(f).
  \end{aligned}
\end{equation}
The component spectra below hold for an \emph{ideal rotor: wind shear, yaw misalignment,
and nacelle tilt are neglected}, so the sectional derivatives $h_{N,i}$ and $h_{T,i}$
are azimuth-independent and each load-transfer coefficient retains only the orders
$s\in\{0,\pm1\}$ of \cref{eq:fourier-coefficients-explicit}; the axial and transverse
components then follow separately.

\subsubsection{Axial Load Spectra}\label{sec:axial-spectra}

For the ideal axial components, only $s=0$ is present, and
\begin{equation}\label{eq:axial-submatrix}
  S_{ad}^{\mathrm{rotor}}(f)
  =
  B^2
  \sum_p
  \sum_{i,j}
  \hat{g}_{i,0}^{(a)}
  K_{ij}^{(pB,pB)}(f)
  \hat{g}_{j,0}^{(d)*},
  \qquad
  a,d\in\{F_x,Q_x\}.
\end{equation}
Consequently,
\begin{equation}\label{eq:thrust-torque-spectra}
  S_{TT}=S_{F_x F_x}^{\mathrm{rotor}},
  \qquad
  S_{TQ}=S_{F_x Q_x}^{\mathrm{rotor}},
  \qquad
  S_{QQ}=S_{Q_x Q_x}^{\mathrm{rotor}}.
\end{equation}

\subsubsection{Transverse Load Spectra}\label{sec:transverse-spectra}

For the transverse components, the circular combinations $F_\pm$ and $Q_\pm$
isolate the azimuthal orders $s = \pm 1$ as established in
\cref{eq:circular-azimuthal-orders}. Evaluating the general CSD formula
\cref{eq:rotor-load-csd} with these single-harmonic sensitivities yields the
circular auto-spectra
\begin{equation}\label{eq:circular-auto-spectra}
  S_{F_\pm F_\pm}(f)
  =
  B^2
  \sum_{p=-\infty}^{\infty}
  \sum_{i,j=1}^{N_r}
  h_{T,i} h_{T,j}
  K_{ij}^{(pB \pm 1, \, pB)}(f),
\end{equation}
with the rotational-sampling kernels $K_{ij}^{(pB \pm 1, \, pB)}(f)$ defined in
\cref{eq:transverse-kernels}.

Because $F_+$ retains only $s = +1$ while $F_-$ retains only $s = -1$ as listed in
\cref{eq:circular-azimuthal-orders}, the product $\hat{g}_{i,s}^{(F_+)}(\hat{g}_{j,s}^{(F_-)})^{\ast}$
vanishes for every order $s \in \mathbb{Z}$. Since the stationary CSD retains only
matched orders $r = s$ of \cref{eq:fourier-orthogonality}, the circular cross-spectrum
vanishes identically:
\begin{equation}\label{eq:circular-cross-zero}
  S_{F_+ F_-}(f) \equiv 0,
  \qquad
  S_{F_- F_+}(f) \equiv 0.
\end{equation}

Applying the linear inverse mapping \cref{eq:cartesian-recovery} in the frequency
domain and substituting \cref{eq:circular-cross-zero} recovers the Cartesian transverse
spectra directly:
\begin{equation}\label{eq:cartesian-spectra-recovery}
  \begin{aligned}
    S_{F_y F_y}(f) = S_{F_z F_z}(f)
    &=
    \frac{1}{4} \left[ S_{F_+ F_+}(f) + S_{F_- F_-}(f) \right], \\[2pt]
    S_{F_y F_z}(f)
    &=
    \frac{\mathrm{i}}{4} \left[ S_{F_+ F_+}(f) - S_{F_- F_-}(f) \right].
  \end{aligned}
\end{equation}
The in-plane auto-spectra $S_{F_y F_y}$ and $S_{F_z F_z}$ are therefore equal,
and their cross-spectrum $S_{F_y F_z}$ is purely imaginary. The moment spectra
$S_{Q_y Q_y}(f)$, $S_{Q_z Q_z}(f)$, and $S_{Q_y Q_z}(f)$ follow from
\cref{eq:circular-auto-spectra,eq:cartesian-spectra-recovery} by replacing
$h_{T,i} \to h_{N,i} r_i$.

\subsubsection{Assembly of the Complete 6-DOF Rotor-Load CSD Matrix}\label{sec:full-csd-matrix}

Because the axial components possess only $s = 0$ while the transverse components
possess only $s = \pm 1$, the product $\hat{g}_{i,s}^{(a)} \hat{g}_{j,s}^{(d)\ast}$
vanishes for any axial--transverse pair with $a \in \{F_x, Q_x\}$ and
$d \in \{F_y, F_z, Q_y, Q_z\}$. Consequently, all axial--transverse cross-spectra
vanish identically:
\begin{equation}\label{eq:axial-transverse-cross-zero}
  S_{F_x F_y} = S_{F_x F_z} = S_{F_x Q_y} = S_{F_x Q_z} = 0,
  \qquad
  S_{Q_x F_y} = S_{Q_x F_z} = S_{Q_x Q_y} = S_{Q_x Q_z} = 0.
\end{equation}
With the component ordering of \cref{eq:rotor-load-vector}, the complete rotor-load
CSD matrix for an ideal rotor decouples into a $2\times 2$ axial block and a $4\times 4$
transverse block:
\begin{equation}\label{eq:hub-load-csd}
  \mathbf{S}_{FF}^{\mathrm{rotor}}(f)=
  \begin{bmatrix}
  S_{F_x F_x} & 0 & 0 & S_{F_x Q_x} & 0 & 0\\
  0 & S_{F_y F_y} & S_{F_y F_z} & 0 & S_{F_y Q_y} & S_{F_y Q_z}\\
  0 & S_{F_z F_y} & S_{F_z F_z} & 0 & S_{F_z Q_y} & S_{F_z Q_z}\\
  S_{Q_x F_x} & 0 & 0 & S_{Q_x Q_x} & 0 & 0\\
  0 & S_{Q_y F_y} & S_{Q_y F_z} & 0 & S_{Q_y Q_y} & S_{Q_y Q_z}\\
  0 & S_{Q_z F_y} & S_{Q_z F_z} & 0 & S_{Q_z Q_y} & S_{Q_z Q_z}
  \end{bmatrix}.
\end{equation}
$\mathbf{S}_{FF}^{\mathrm{rotor}}(f)$ is Hermitian and positive semidefinite.


\section{Numerical Implementation}\label{sec:numerical}


\Cref{eq:rotor-load-csd} expresses the rotor-load CSD as a sum over the
blade-passing family $p$ and the sensitivity order $s$ of radial double sums over
station pairs. Its evaluation requires the coherence harmonic $F_{pB+s}$ on the
$N_r\times N_r$ station pairs and the transfer coefficients $\hat{g}_{i,s}^{(a)}$
at the $N_r$ stations. Three uniform grids discretize the continuous variables,
listed in \cref{tab:discretization-grids}.

\begin{table}[H]
  \centering
  \caption{Uniform grids of the discrete evaluation.}
  \label{tab:discretization-grids}
  \begin{tabular}{@{}lll@{}}
    \toprule
    Grid & Nodes & Quantity evaluated \\
    \midrule
    Azimuthal quadrature
      & $\theta_\mu=\dfrac{\pi\mu}{N_\theta}$,\quad $\mu=0,\ldots,N_\theta$
      & $F_n$ (\cref{eq:coherence-harmonic-quadrature}) \\
    Azimuthal load sampling
      & $\psi_\kappa=\dfrac{2\pi(\kappa-1)}{N_\psi}$,\quad
        $\kappa=1,\ldots,N_\psi$
      & $\hat{g}_{i,s}^{(a)}$ (\cref{eq:sensitivity-dft}) \\
    Output frequency
      & $f_\ell=f_1+(\ell-1)\Delta f$,\quad $\ell=1,\ldots,N_f$
      & $\mathbf{S}(f_\ell)$ (\cref{eq:csd-assembly}) \\
    \bottomrule
  \end{tabular}
\end{table}

$N_\theta$ counts intervals on the half period $[0,\pi]$, the full revolution
carrying $2N_\theta$ mirror-symmetric samples \cref{eq:separation-even}; $f_1$,
$\Delta f$, and $N_f$ are the lowest reported frequency, the frequency spacing,
and the number of bins.

The derivation is carried out with two-sided densities, whereas the reported
spectra are one-sided,
\begin{equation}\label{eq:one-sided-reporting}
  S_{ad}^{1s}(f)=2S_{ad}(f),
  \qquad f\ge0,
\end{equation}
which folds the negative-frequency half-axis onto $f\ge0$. Applied to the even
two-sided Kaimal representation \cref{eq:kaimal-two-sided}, it recovers the
one-sided IEC ordinate \cref{eq:kaimal}; on the diagonal it gives
$\int_0^\infty S_{aa}^{1s}(f)\,\mathrm{d}f=R_{aa}(0)$.

The component spectra of \cref{sec:hub-load-csd-matrix} and the block structure of
\cref{eq:hub-load-csd} hold for the ideal rotor, whose azimuth-independent
sectional derivatives restrict the transfer coefficients to $s\in\{0,\pm1\}$ and
decouple the axial and transverse components. That restriction is not used below:
the discretization accepts general azimuth-dependent transfer coefficients, so
every retained order contributes and the two groups couple through the matched
pairs $(p,s)$.


\subsection{Coherence Harmonic Matrix}


The coherence harmonic corresponds to
the cosine integral \cref{eq:Fn} of the
IEC coherence \cref{eq:iec-coherence},
with its discrete representation given by the DFT
over a full revolution.
Since $d_{ij}(\theta)$ is $2\pi$-periodic
\cref{eq:separation-periodic} and symmetric about both $\theta=0$
and $\theta=\pi$ \cref{eq:separation-even},
the coefficients satisfy $F_{-n} = F_n$.
Consistent with the defined one-sided spectrum,
the negative orders are folded onto $n \ge 0$,
and the summation reduces to the half-period $[0, \pi]$:
\begin{equation}\label{eq:coherence-harmonic-quadrature}
  F_n(f;r_i,r_j)
  \approx
  \frac{1}{N_\theta}
  \sum_{\mu=0}^{N_\theta}
  w_\mu\,
  \gamma\!\left(f,d_{ij}(\theta_\mu)\right)\cos(n\theta_\mu),
  \qquad
  \theta_\mu=\frac{\pi\mu}{N_\theta},
  \qquad
  w_0=w_{N_\theta}=\tfrac{1}{2},
\end{equation}
with $w_\mu=1$ for $1\le\mu\le N_\theta-1$; the half weights count the two
symmetry points once. Equation \cref{eq:coherence-harmonic-quadrature} reproduces
\cref{eq:Fn} up to the aliased orders, $F_n\rightarrow\sum_m F_{n+2mN_\theta}$,
whose contribution is exponentially small in $N_\theta$ for the analytic coherence
\cref{eq:iec-coherence}. For fixed $f$ and $n$, the harmonic on the station pairs
forms the real symmetric matrix
\begin{equation}\label{eq:coherence-harmonic-matrix}
  \mathbf{F}_n(f)
  =
  \begin{bmatrix}
    F_n(f;r_1,r_1) & \cdots & F_n(f;r_1,r_{N_r})\\
    \vdots & \ddots & \vdots\\
    F_n(f;r_{N_r},r_1) & \cdots & F_n(f;r_{N_r},r_{N_r})
  \end{bmatrix}
  \in\mathbb{R}^{N_r\times N_r}.
\end{equation}
Realness follows from the cosine form \cref{eq:Fn}, and symmetry from the
interchange of $r_i$ and $r_j$ in \cref{eq:separation}, so
$\mathbf{F}_n(f)=\mathbf{F}_n^{\mathrm{T}}(f)$.


\subsection{Modulated Kernel Matrix}


Each blade-passing family $p$ shifts the incident turbulence spectrum to the
fixed-frame frequency $\nu$ of \cref{eq:q-and-n-boxed}. Both $\mathbf{F}_n$ and
$S_u$ depend on $\nu$ only through its magnitude
\begin{equation}\label{eq:shifted-frequency-magnitude}
  \lvert\nu\rvert=\lvert f-pBf_{\mathrm{rot}}\rvert ,
\end{equation}
so for each output frequency $f$ the magnitude is formed once, and all $N_r^2$
entries of the kernel follow from the coherence harmonic
\cref{eq:coherence-harmonic-matrix} and the wind spectrum
\cref{eq:kaimal-two-sided} at that argument. Scaling the coherence harmonic by
the two-sided wind auto-spectral density $S_u(\nu)$ yields the
$N_r\times N_r$ radial kernel matrix
\begin{equation}\label{eq:kernel-matrix}
  \begin{aligned}
    \mathbf{K}^{(n,q)}(f)
    &=
    \mathbf{F}_n(\nu)\,S_u(\nu)\\
    &=
    \begin{bmatrix}
      F_n(\nu;r_1,r_1)S_u(\nu) & \cdots & F_n(\nu;r_1,r_{N_r})S_u(\nu)\\
      \vdots & \ddots & \vdots\\
      F_n(\nu;r_{N_r},r_1)S_u(\nu) & \cdots & F_n(\nu;r_{N_r},r_{N_r})S_u(\nu)
    \end{bmatrix}.
  \end{aligned}
\end{equation}
with $n=pB+s$ and $q=pB$. The family sum is truncated symmetrically at
$\lvert p\rvert\le p_{\max}$.


\subsection{Azimuthal Transform of the Load Sensitivities}\label{sec:sensitivity-transform}


The transfer coefficients $g_i^{(a)}(\psi)$ of \cref{eq:sensitivities-def} are
composed from the sectional sensitivities $h_{N,i}$ and $h_{T,i}$ introduced in
\cref{eq:sectional-linearization}.
At each azimuth they are obtained by the centred difference
\begin{align}
  h_{N,i}(\psi_b)
  &=
  \frac{N_{ib}\!\left(u_{ib}+\varepsilon;\psi_b\right)-N_{ib}\!\left(u_{ib}-\varepsilon;\psi_b\right)}
       {2\varepsilon},
  \nonumber\\
  h_{T,i}(\psi_b)
  &=
  \frac{T_{ib}\!\left(u_{ib}+\varepsilon;\psi_b\right)-T_{ib}\!\left(u_{ib}-\varepsilon;\psi_b\right)}
       {2\varepsilon},
  \label{eq:centered-difference}
\end{align}
with $u_{ib}$ the operating inflow at station $i$ of blade $b$ and $\varepsilon$ the local inflow
speed perturbation, set to $\varepsilon=10^{-6}$~m/s.

The azimuthal dependence of the transfer coefficients \cref{eq:delta-F} is sampled
on the uniform grid
\begin{equation}\label{eq:load-sampling-grid}
  \psi_\kappa=\frac{2\pi(\kappa-1)}{N_\psi},
  \qquad
  \kappa=1,\ldots,N_\psi,
\end{equation}
and transformed by the DFT of the sampled coefficients, which reproduces the
defining integral \cref{eq:sensitivity-expansion} up to aliasing,
\begin{equation}\label{eq:sensitivity-dft}
  \hat{g}_{i,s}^{(a)}
  =
  \frac{1}{N_\psi}
  \sum_{\kappa=1}^{N_\psi}
  g_i^{(a)}(\psi_\kappa)\,
  \ee^{-\mathrm{i}2\pi s(\kappa-1)/N_\psi},
  \qquad
  \lvert s\rvert\le s_{\max}=\left\lfloor N_\psi/2\right\rfloor-1 .
\end{equation}
The bound $s_{\max}=\lfloor N_\psi/2\rfloor-1$ keeps the retained band strictly
below the aliasing limit $N_\psi/2$ and removes the terminal order
$s=N_\psi/2$, at which the coefficients of $+s$ and $-s$ coincide, when
$N_\psi$ is even. A shift of the grid origin by $\psi_0$ multiplies every
coefficient by $\ee^{\mathrm{i}s\psi_0}$, a factor that cancels in the products
$\hat{g}_{i,s}^{(a)}\hat{g}_{j,s}^{(d)\ast}$ of \cref{eq:rotor-load-csd}; the
grid origin is therefore set to zero.

The ideal-rotor coefficients \cref{eq:fourier-coefficients-explicit} occupy only
$s\in\{0,\pm1\}$, while azimuth-dependent aerodynamic inputs spread the harmonic
content over higher orders. For each order $s$, these coefficients form the
transfer matrix
\begin{equation}\label{eq:transfer-matrix}
  \mathbf{G}_s
  =
  \begin{bmatrix}
    \hat{g}_{1,s}^{(F_x)} & \cdots & \hat{g}_{N_r,s}^{(F_x)}\\
    \vdots & \ddots & \vdots\\
    \hat{g}_{1,s}^{(Q_z)} & \cdots & \hat{g}_{N_r,s}^{(Q_z)}
  \end{bmatrix}
  \in\mathbb{C}^{6\times N_r},
  \qquad
  \left(\mathbf{G}_s\right)_{ai}=\hat{g}_{i,s}^{(a)},
\end{equation}
%


\subsection{Assembly of the Load CSD}


Substituting \cref{eq:kernel-matrix} and \cref{eq:transfer-matrix} into the general
expression \cref{eq:rotor-load-csd}, the radial double sum of a retained order
pair $(p,s)$ reduces to the $6\times6$ matrix
\begin{equation}\label{eq:csd-order-term}
  \mathbf{S}^{(p,s)}(f)
  \coloneqq
  \mathbf{G}_s\,\mathbf{K}^{(pB+s,\,pB)}(f)\,\mathbf{G}_s^{\mathrm{H}},
\end{equation}
whose entries are, component by component,
\begin{equation}\label{eq:csd-order-entry}
  S_{ad}^{(p,s)}(f)
  =
  \sum_{i=1}^{N_r}\sum_{j=1}^{N_r}
  \hat{g}_{i,s}^{(a)}
  K_{ij}^{(pB+s,\,pB)}(f)
  \hat{g}_{j,s}^{(d)\ast},
  \qquad
  a,d\in\mathcal{L}.
\end{equation}
The blade factor $B^{2}$ of \cref{eq:stationary-covariance-blade-summed} multiplies
the family sum, so the sums over $s$ and $p$ give
\begin{equation}\label{eq:csd-assembly}
  \mathbf{S}(f)
  =
  B^{2}
  \sum_{p=-p_{\max}}^{p_{\max}}
  \sum_{s\in\mathcal{S}}
  \mathbf{G}_s\,\mathbf{K}^{(pB+s,\,pB)}(f)\,\mathbf{G}_s^{\mathrm{H}} .
\end{equation}
Truncation at $\lvert p\rvert=p_{\max}$ is admissible because $S_u(\nu)$ decays
as $\lvert\nu\rvert^{-5/3}$ and $F_n(\nu)$ decays exponentially in $\lvert n\rvert$
for the analytic coherence \cref{eq:iec-coherence}.

Every summand of \cref{eq:csd-assembly} is Hermitian in exact arithmetic:
$K_{ij}=K_{ji}$ is real, and \cref{eq:csd-order-entry} gives
$S_{ad}^{(p,s)}=\left(S_{da}^{(p,s)}\right)^{\ast}$. The symmetry is imposed after
summation,
\begin{equation}\label{eq:csd-hermitian-projection}
  \mathbf{S}(f)
  \leftarrow
  \tfrac{1}{2}\left(\mathbf{S}(f)+\mathbf{S}^{\mathrm{H}}(f)\right),
\end{equation}
which removes the antisymmetric part produced by round-off. Positive
semidefiniteness is not implied by \cref{eq:csd-assembly} once $p$ and
$\mathcal{S}$ are truncated, and is therefore verified at each frequency before
$\mathbf{S}(f)$ is factorized.

\section{Validation}\label{sec:validation}


This section validates the rotor-load CSD of \cref{eq:hub-load-csd} against simulation
results of a floating offshore wind turbine (FOWT), which is built on the NLR 5MW
reference wind turbine (formerly the NREL 5MW turbine) and is adopted here as the
validation object. The main parameters of the turbine are listed
in \cref{tab:nlr5mw} and
the nine load cases in \cref{tab:validation-cases}.

\begin{table}[H]
  \centering
  \caption{Main parameters of the NLR 5-MW reference wind turbine~\cite{NREL-5MW}.}
  \label{tab:nlr5mw}
  {\small
  \setlength{\tabcolsep}{6pt}%
  \begin{tabular}{@{}ll@{}}
    \toprule
    Parameter & Value \\
    \midrule
    Configuration          & 3 blades \\
    Control                & Variable speed, collective pitch \\
    Rotor diameter         & 126~m \\
    Hub height             & 90~m \\
    Rated wind speed       & 11.4~m/s \\
    Cut-in, rated rotor speed & 6.9~rpm, 12.1~rpm \\
    Overhang               & 5~m \\
    Shaft tilt             & $5^{\circ}$ \\
    Precone                & $2.5^{\circ}$ \\
    \bottomrule
  \end{tabular}}
\end{table}

The reference results are computed using OpenFAST, an open-source multi-physics
simulation code for wind turbines~\cite{OpenFAST}.
All cases use the IEC Kaimal spectrum with turbulence class II-B under the normal
turbulence model (NTM)~\cite{IEC-2019}, that is, wind class II and
turbulence category B with $I_{\mathrm{ref}}=0.14$ in \cref{eq:sigma1}.
The hydro, control, and aeroelastic (ElastoDyn) modules are switched off, so the FOWT
behaves as a rigid body and no dynamic reaction to the aerodynamic forces is computed;
the blade motion is then fixed by the rotor speed alone, and the load fluctuation
originates in the turbulent inflow only.
For each of the nine inflow speeds, the rotor speed and the blade pitch angle take the
steady-state values of that inflow speed, listed in the table.

\begin{table}[H]
  \centering
  \caption{Validation load cases.}
  \label{tab:validation-cases}
  {\small
  \setlength{\tabcolsep}{5pt}%
  \begin{tabular}{@{}ccccccc@{}}
    \toprule
    $U_\infty$ (m/s) & Turbulence & Hydro & Control & Aeroelasticity
      & Rotor speed (rpm) & Pitch (deg) \\
    \midrule
    6  & IEC Kaimal IIB NTM & \-- & \-- & \-- & 7.94  & 0.00  \\
    8  & IEC Kaimal IIB NTM & \-- & \-- & \-- & 9.16  & 0.00  \\
    10 & IEC Kaimal IIB NTM & \-- & \-- & \-- & 11.43 & 0.00  \\
    11 & IEC Kaimal IIB NTM & \-- & \-- & \-- & 11.89 & 0.00  \\
    12 & IEC Kaimal IIB NTM & \-- & \-- & \-- & 12.10 & 3.83  \\
    14 & IEC Kaimal IIB NTM & \-- & \-- & \-- & 12.10 & 8.70  \\
    16 & IEC Kaimal IIB NTM & \-- & \-- & \-- & 12.10 & 12.06 \\
    18 & IEC Kaimal IIB NTM & \-- & \-- & \-- & 12.10 & 14.92 \\
    20 & IEC Kaimal IIB NTM & \-- & \-- & \-- & 12.10 & 17.47 \\
    \bottomrule
  \end{tabular}}
\end{table}

The force and moment channels \texttt{RtAeroF$\xi$i} and \texttt{RtAeroM$\xi$i}
($\xi=x,y,z$, the trailing letter marking the inertial frame) output by OpenFAST are
used for the comparison, transformed from the inertial frame to the non-rotating hub
frame through the $-5^{\circ}$ shaft tilt of the rotor axis; the load-transfer
coefficients of the present theory are obtained from the blade-element momentum code
CCBlade~\cite{CCBlade1,CCBlade2}.
Each OpenFAST simulation case consists of a 3600 s TurbSim inflow record and
3600 s of aerodynamic simulation.
The OpenFAST time series, sampled at $\Delta T=0.0125$~s, is converted
into PSDs by Welch's method with $2^{14}$ samples per segment, which gives a
frequency resolution of $4.9\times10^{-3}$~Hz. 

\begin{figure}[H]
  \centering
  \includegraphics[width=0.95\textwidth]{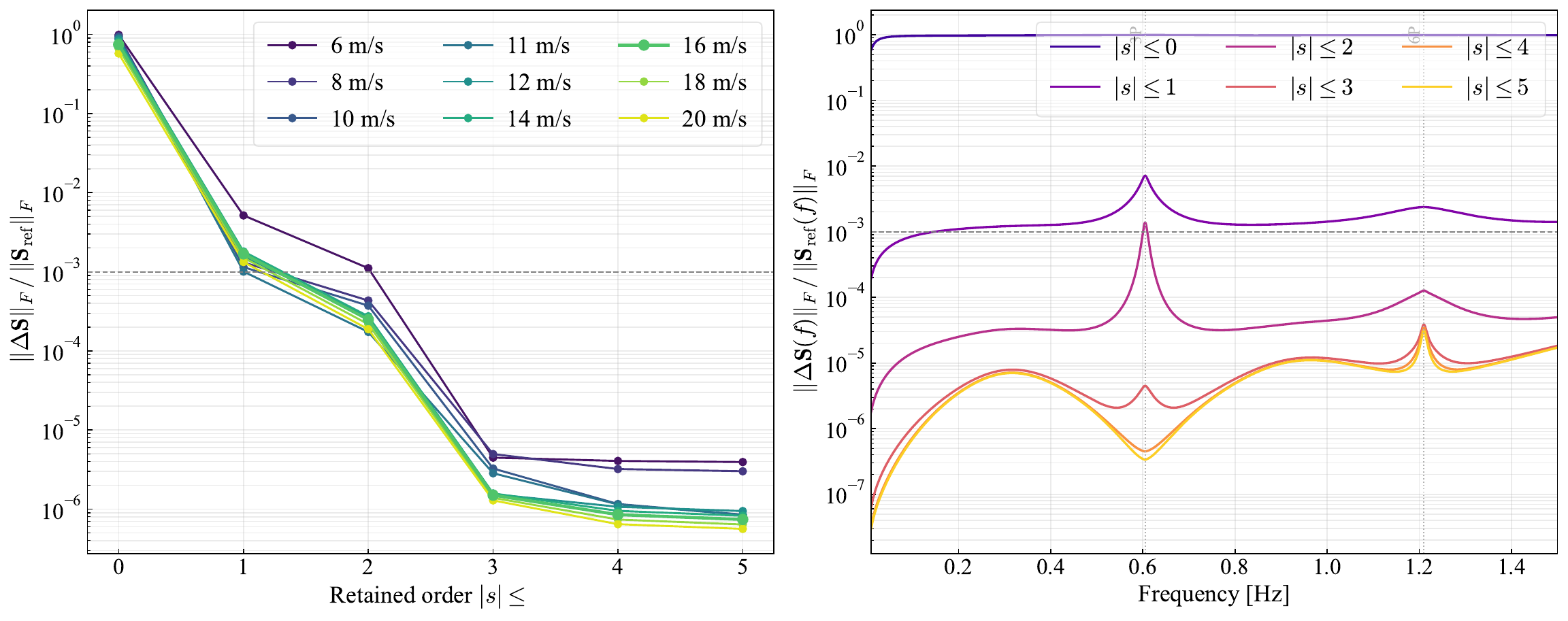}
  \caption{Convergence of the assembled rotor-load CSD with the retained azimuthal
  order $s$ of the load-transfer coefficients. The reference
  $\mathbf{S}_{\mathrm{ref}}$ is assembled with the full band
  $\lvert s\rvert\le s_{\max}=17$.}
  \label{fig:s-converge}
\end{figure}

Two sums of \cref{eq:csd-assembly} are truncated: the blade-passing family $p$ and
the azimuthal order $s$ of the load-transfer coefficients.
Harmonics above the third blade-passing family are not of interest here, so the
family sum keeps $\lvert p\rvert\le 3$.
The order $s$ is set by a convergence test.
In the steady BEMT calculation done with CCBlade, the transfer coefficients are
sampled at $N_\psi=36$ azimuths, and the band bound \cref{eq:sensitivity-dft} then
gives $s_{\max}=17$.
The CSD assembled with the full band $\lvert s\rvert\le s_{\max}$ is taken as the
reference $\mathbf{S}_{\mathrm{ref}}$, and repeating the assembly with a smaller
band $\lvert s\rvert\le s_{\mathrm{c}}$ gives \cref{fig:s-converge}, where the
relative error
$\lVert\mathbf{S}-\mathbf{S}_{\mathrm{ref}}\rVert_{\mathrm{F}}/
\lVert\mathbf{S}_{\mathrm{ref}}\rVert_{\mathrm{F}}$ keeps falling as orders are
added and stays below $10^{-5}$ from $\lvert s\rvert\le4$ on, so this band is kept
in the rest of the validation.

\begin{figure}[H]
  \centering
  \begin{subfigure}[b]{0.49\textwidth}
    \centering
    \includegraphics[width=\textwidth]{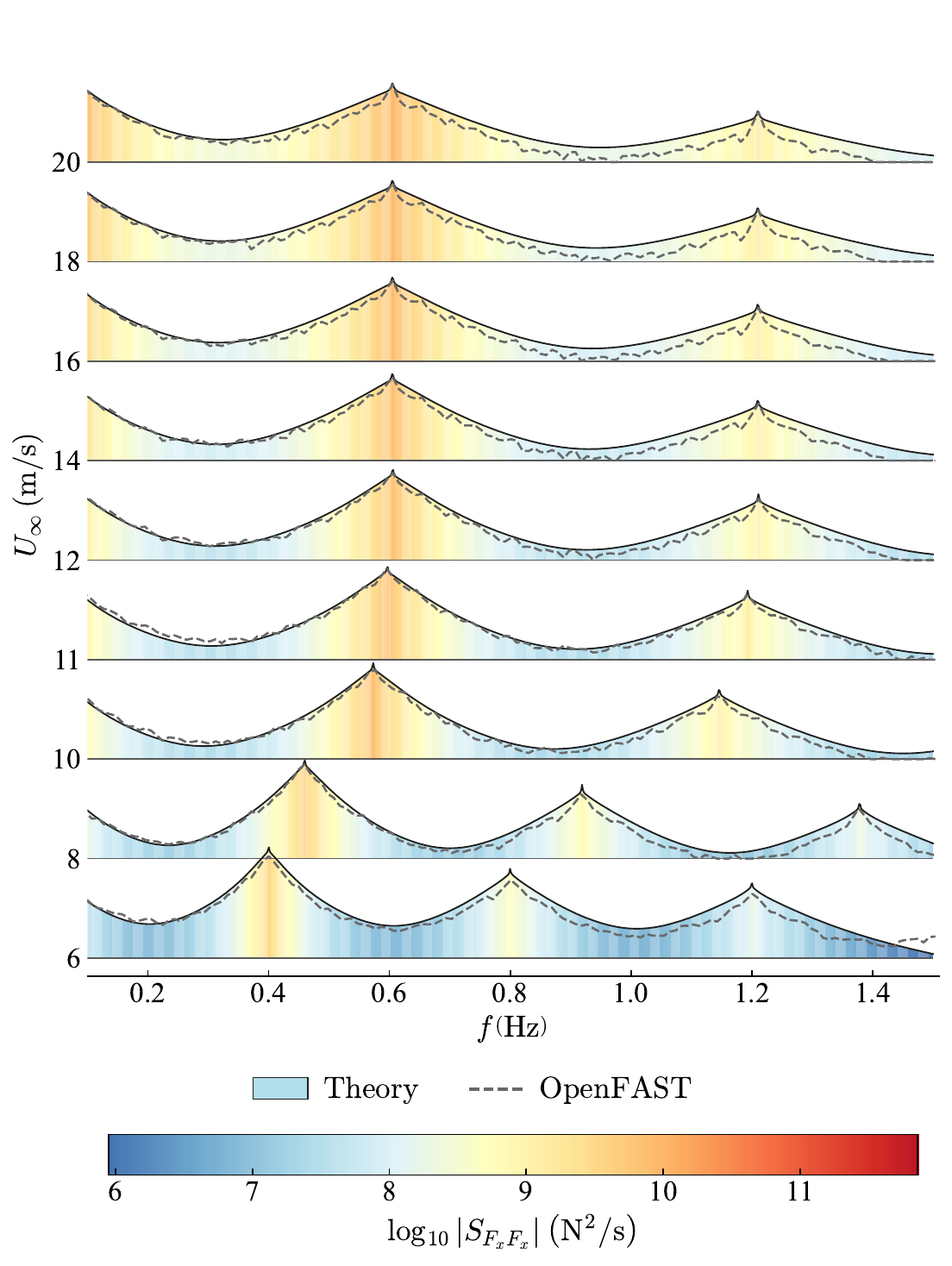}
    \caption{Thrust $S_{F_x F_x}(f)$.}
    \label{fig:val-axial-fx}
  \end{subfigure}\hfill
  \begin{subfigure}[b]{0.49\textwidth}
    \centering
    \includegraphics[width=\textwidth]{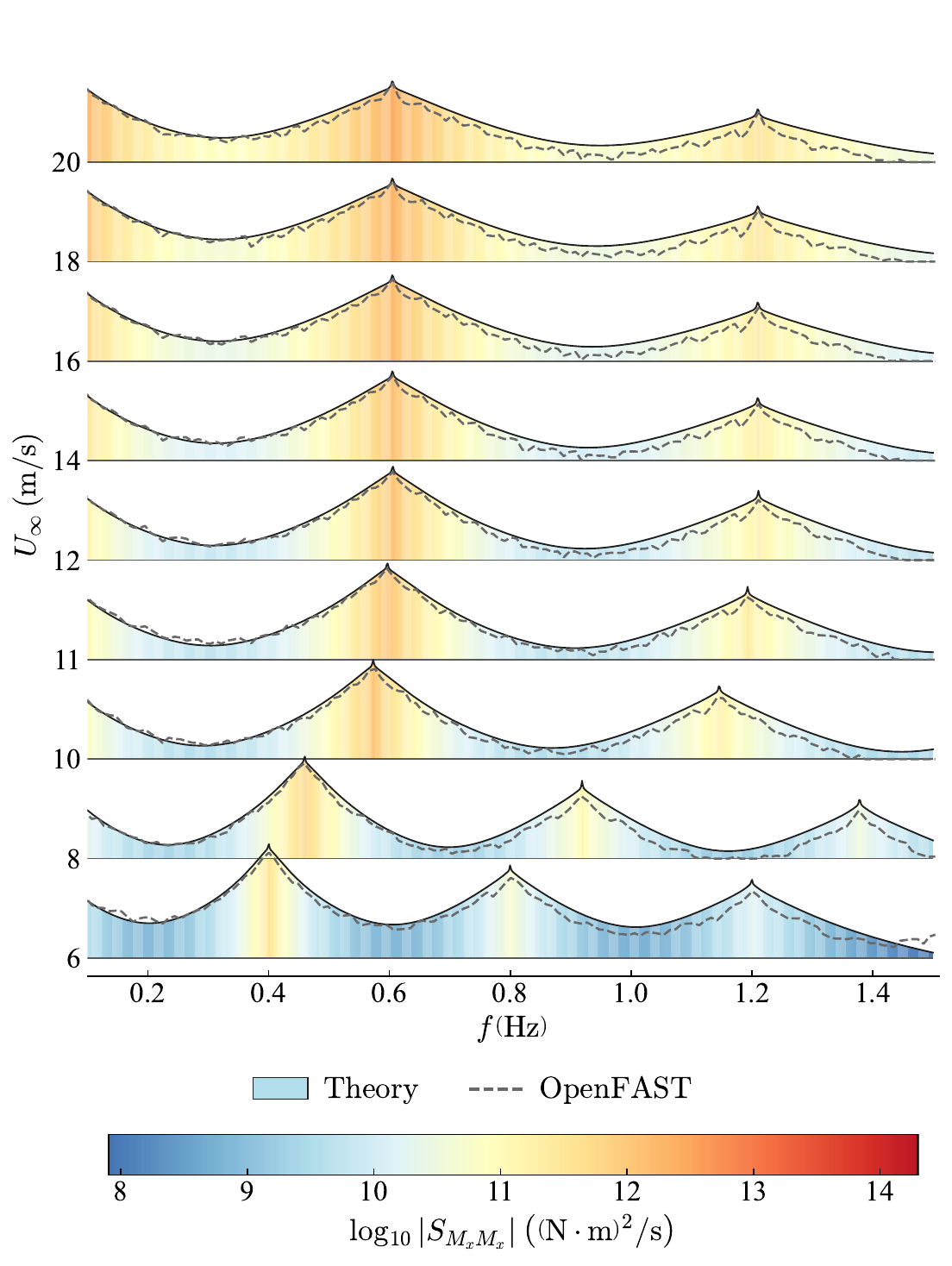}
    \caption{Shaft torque $S_{Q_x Q_x}(f)$.}
    \label{fig:val-axial-qx}
  \end{subfigure}
  \caption{Axial load cross-spectral densities as a function of the observed
  frequency $f$, stacked over the nine wind speeds of
  \cref{tab:validation-cases} ($U_\infty$ denotes the hub wind speed
  $U_{\mathrm{hub}}$): (a) thrust $S_{F_x F_x}(f)$; (b) shaft torque
  $S_{Q_x Q_x}(f)$. In each panel the filled ridge is the present theory, the
  dashed line is the OpenFAST result, and the colour gives the magnitude
  $\log_{10}\lvert S_{aa}\rvert$ on the scale below the axes.}
  \label{fig:val-axial}
\end{figure}

\begin{figure}[H]
  \centering
  \begin{subfigure}[b]{0.47\textwidth}
    \centering
    \includegraphics[width=\textwidth]{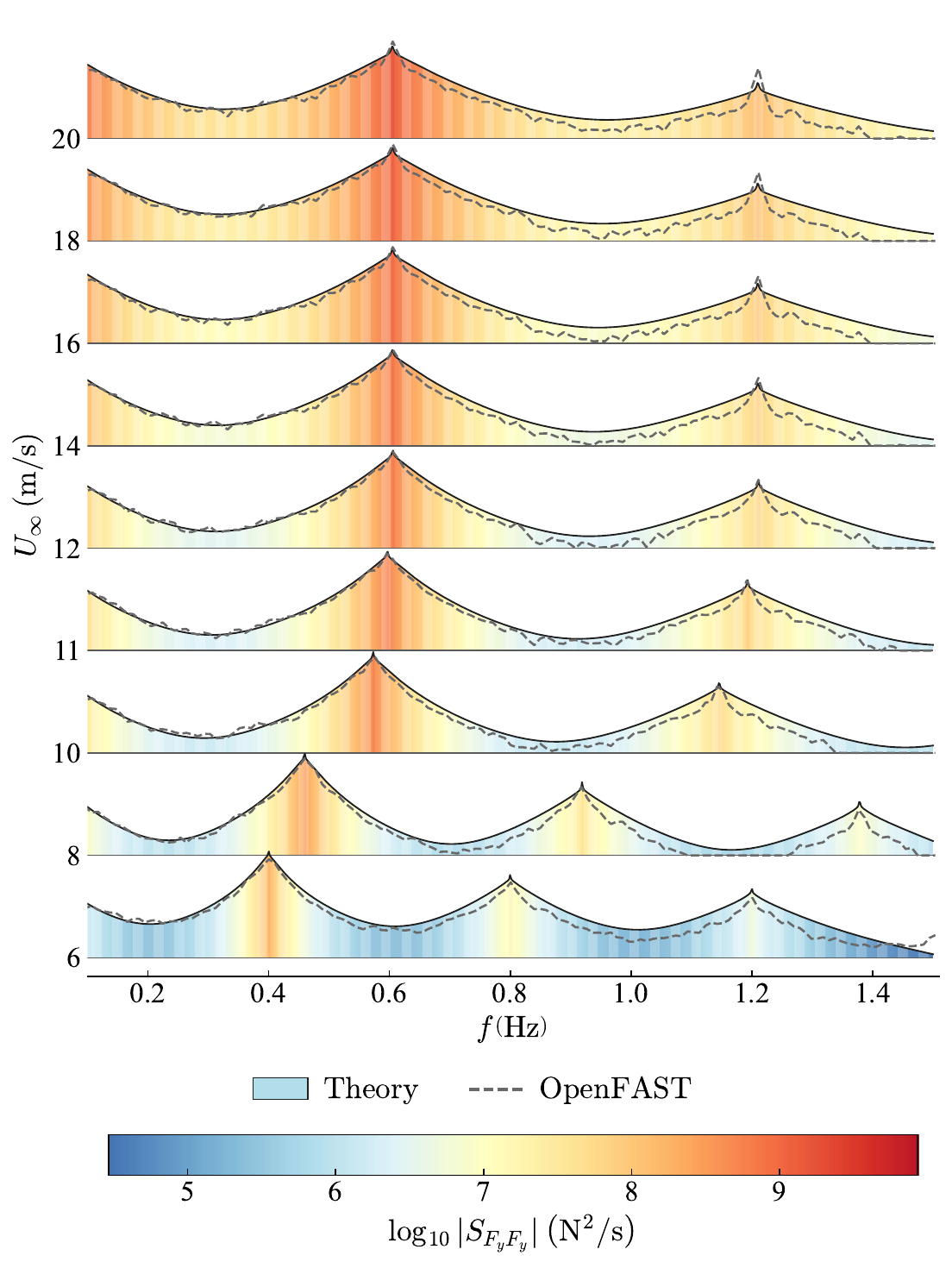}
    \caption{Lateral force $S_{F_y F_y}(f)$.}
    \label{fig:val-transverse-fy}
  \end{subfigure}\hfill
  \begin{subfigure}[b]{0.47\textwidth}
    \centering
    \includegraphics[width=\textwidth]{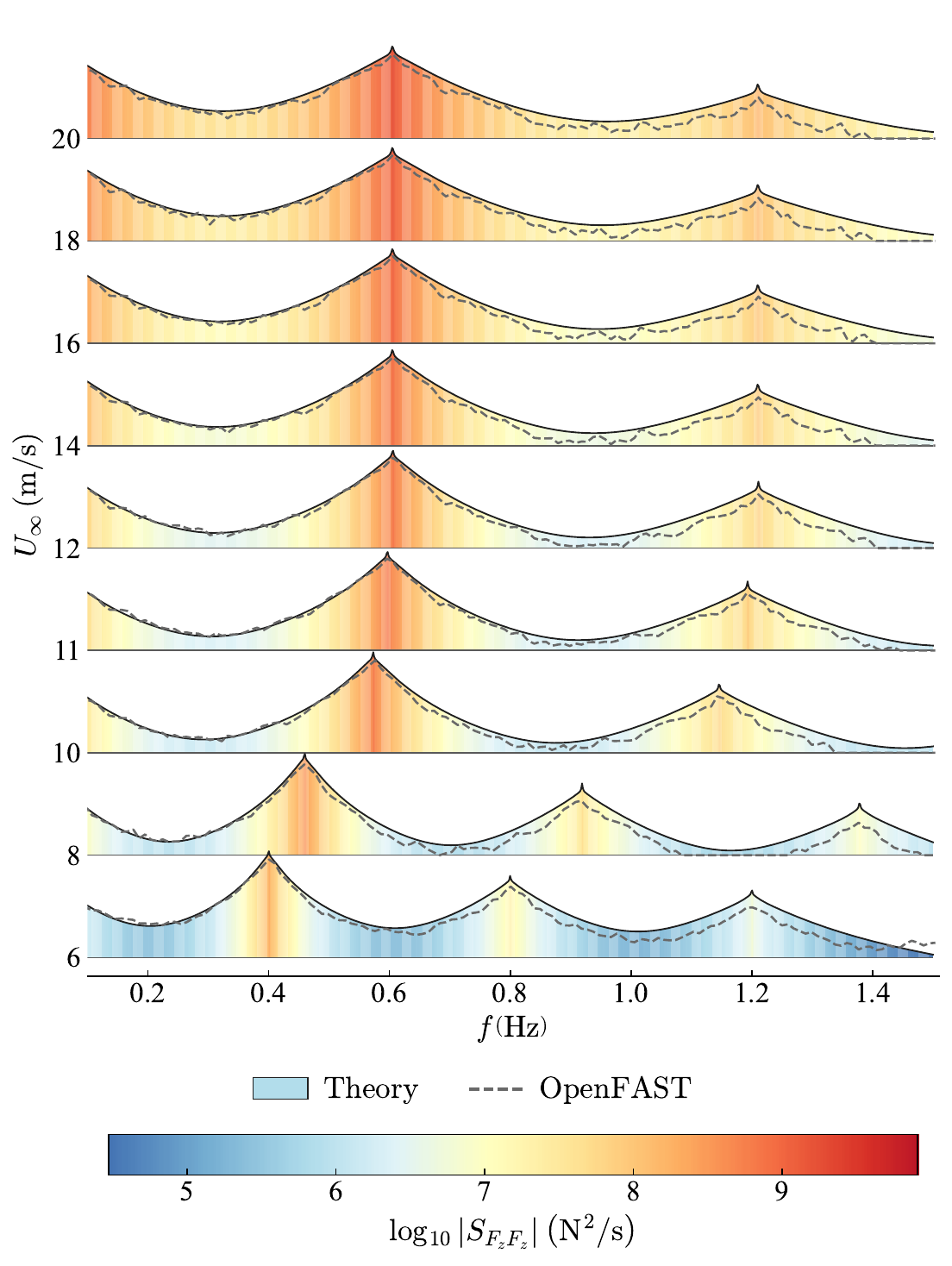}
    \caption{Vertical force $S_{F_z F_z}(f)$.}
    \label{fig:val-transverse-fz}
  \end{subfigure}

  \vspace{1ex}
  \begin{subfigure}[b]{0.47\textwidth}
    \centering
    \includegraphics[width=\textwidth]{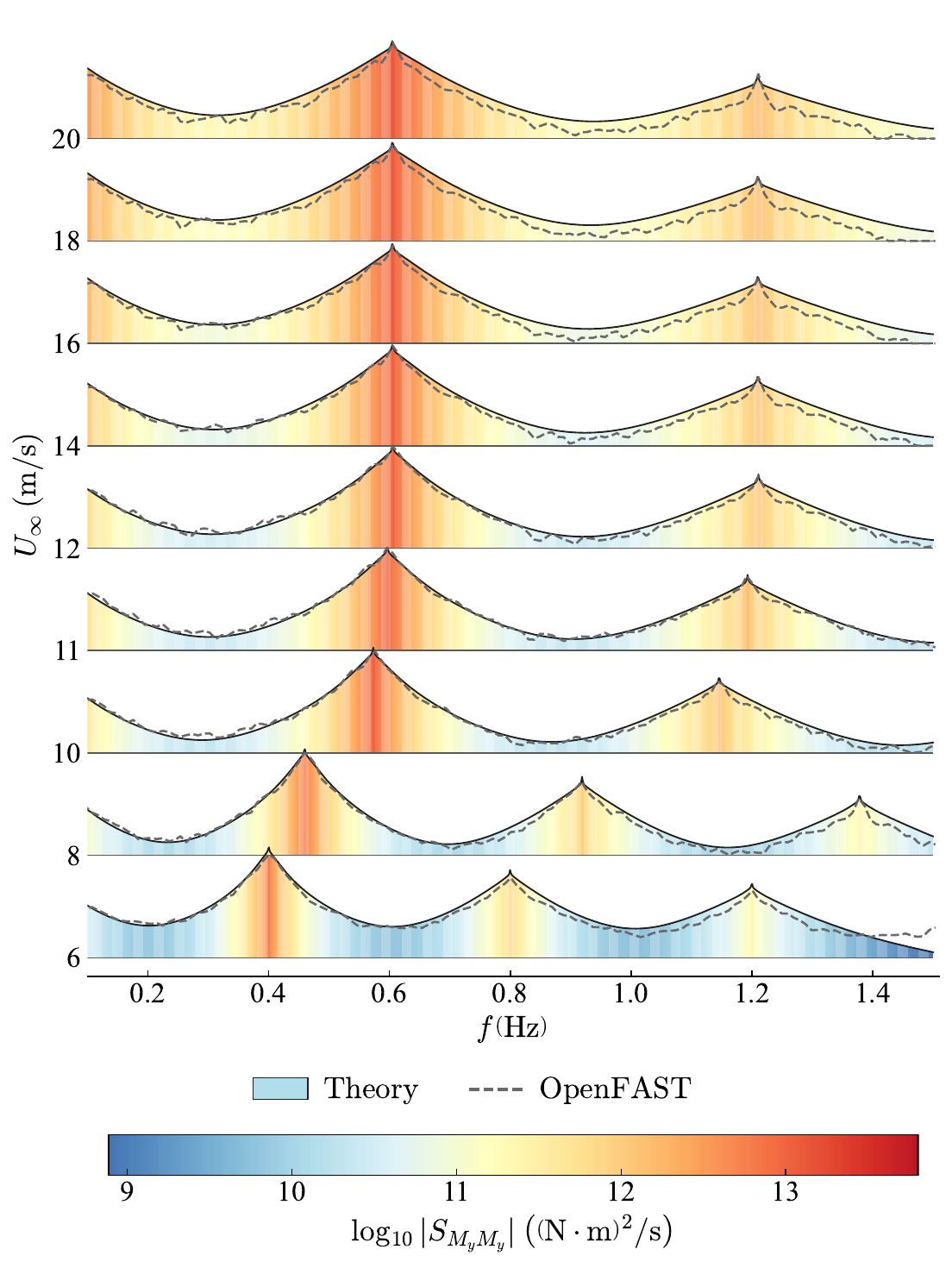}
    \caption{Nodding moment $S_{Q_y Q_y}(f)$.}
    \label{fig:val-transverse-qy}
  \end{subfigure}\hfill
  \begin{subfigure}[b]{0.47\textwidth}
    \centering
    \includegraphics[width=\textwidth]{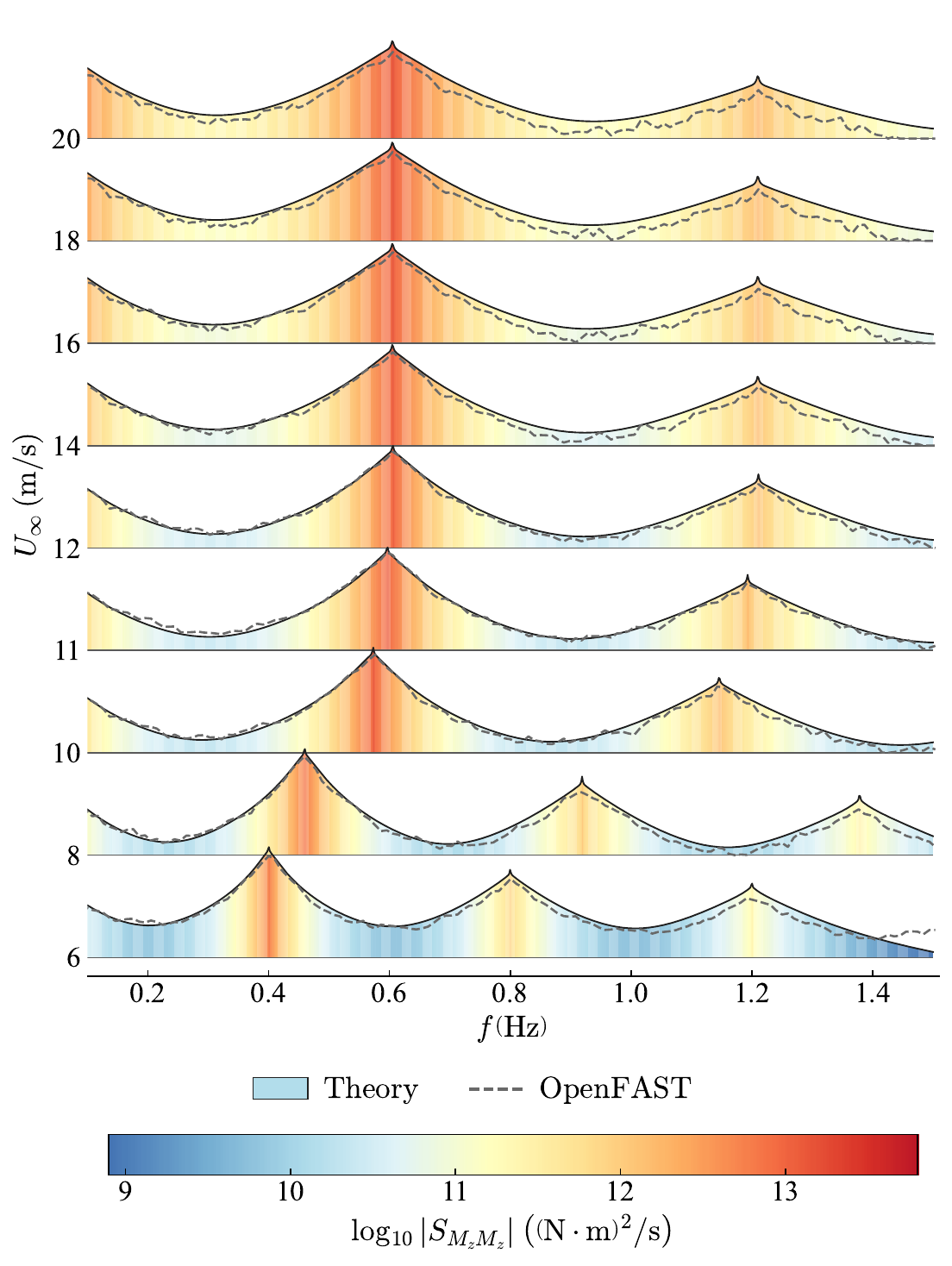}
    \caption{Yawing moment $S_{Q_z Q_z}(f)$.}
    \label{fig:val-transverse-qz}
  \end{subfigure}
  \caption{Transverse load cross-spectral densities at the same nine wind
  speeds: (a) lateral force $S_{F_y F_y}(f)$; (b) vertical force
  $S_{F_z F_z}(f)$; (c) nodding moment $S_{Q_y Q_y}(f)$; (d) yawing moment
  $S_{Q_z Q_z}(f)$. Presentation and colour scale as in \cref{fig:val-axial}.}
  \label{fig:val-transverse}
\end{figure}

\Cref{fig:val-axial} and \cref{fig:val-transverse} present the six auto-spectral
densities as coloured ridge plots stacked over the nine wind speeds of
\cref{tab:validation-cases}: the filled ridge is the present theory, whose height and
colour give $\log_{10}\lvert S_{aa}\rvert$ at each frequency $f$ and hub wind speed,
and the dashed line is the OpenFAST result at the same operating point.
Because both descriptions share the same axes, they are compared directly at each
frequency.
They agree closely below 1~Hz in all six components, in both magnitude and shape.
Above 1~Hz the present theory exceeds the OpenFAST PSD around the second and
third blade-passing families $p=2$ and $p=3$ of \cref{eq:rotor-load-csd}, that is,
around the 6P and 9P harmonics of the three-bladed rotor, which for the cases with
rotor speed near 11.5~rpm lie at 1.15 and 1.72~Hz.

\section*{Acknowledgement}



Large language models were applied in the theory derivation of this manuscript.
Within the derivation, the formulation was informed by DeepSeek Flash 4.1 and
corrected by ChatGPT 5.6 Sol and DeepSeek Flash 4.1.


\bibliographystyle{unsrtnat}
\bibliography{references}
\end{document}